\documentclass[preprint,12pt,authoryear]{elsarticle}

\usepackage{amssymb}
\usepackage{url}

\usepackage{breakurl}
\usepackage{subcaption}
\usepackage{graphicx}
\usepackage{pdflscape}

\journal{Array}

\begin{document}

\begin{frontmatter}



\title{Privacy Impact Assessments in the Wild: A Scoping Review}


\author[inst1]{Leonardo Horn Iwaya}
\affiliation[inst1]{organization={Department of Mathematics and Computer Science, Karlstad University},
            addressline={Universitetsgatan 2}, 
            city={Karlstad},
            postcode={651 88}, 
            state={Värmland},
            country={Sweden}}

\author[inst2]{Ala Sarah Alaqra}
\affiliation[inst2]{organization={Information Systems, Karlstad Business School, Karlstad University},
            addressline={Universitetsgatan 2}, 
            city={Karlstad},
            postcode={651 88}, 
            state={Värmland},
            country={Sweden}}

\author[inst3]{Marit Hansen}
\affiliation[inst3]{organization={Unabhängiges Landeszentrum für Datenschutz Schleswig-Holstein},
            addressline={Holstenstraße 98}, 
            city={Kiel},
            postcode={24103}, 
            state={Schleswig-Holstein},
            country={Germany}}

\author[inst1,inst4]{Simone Fischer-Hübner}
\affiliation[inst4]{organization={Department of Computer Science and Engineering
Chalmers University of Technology},
            addressline={Chalmersplatsen 4}, 
            city={Gothenburg},
            postcode={412 96}, 
            state={Västra Götaland},
            country={Sweden}}

\begin{abstract}
Privacy Impact Assessments (PIAs) offer a process for assessing the privacy impacts of a project or system. As a privacy engineering strategy, they are one of the main approaches to privacy by design, supporting the early identification of threats and controls. However, there is still a shortage of empirical evidence on their use and proven effectiveness in practice. To better understand the current literature and research, this paper provides a comprehensive Scoping Review (ScR) on the topic of PIAs ``in the wild,'' following the well-established Preferred Reporting Items for Systematic reviews and Meta-Analyses (PRISMA) guidelines. This ScR includes 45 studies, providing an extensive synthesis of the existing body of knowledge, classifying types of research and publications, appraising the methodological quality of primary research, and summarising the positive and negative aspects of PIAs in practice, as reported by those studies. This ScR also identifies significant research gaps (e.g., evidence gaps from contradictory results and methodological gaps from research design deficiencies), future research pathways, and implications for researchers, practitioners, and policymakers developing and using PIA frameworks. As we conclude, there is still a significant need for more primary research on the topic, both qualitative and quantitative. A critical appraisal of qualitative studies revealed deficiencies in the methodological quality, and only four quantitative studies were identified, suggesting that current primary research remains incipient. Nonetheless, PIAs can be regarded as a prominent sub-area in the broader field of empirical privacy engineering, in which further scientific research to support existing practices is needed.
\end{abstract}


\end{frontmatter}


\section{Introduction}
\label{sec:introduction}


A Privacy Impact Assessment (PIA) has been defined by Wright~\citep{wright2012state} as a \textit{``methodology for assessing the impacts on privacy of a project, policy, programme, service, product or other initiative and, in consultation with stakeholders, for taking remedial actions as necessary in order to avoid or minimise negative impacts.''} It is a process that should begin at the earliest possible stage of a project and should continue even after the project has been implemented \citep{wright2012state}.
PIAs emerged in the last century as a specific form of Technology Assessment and have become progressively more commonly used in practice from the mid-1990s onwards \citep{CLARKE2009123}.

A related methodology is a Data Protection Impact Assessment (DPIA), which has been incorporated into the EU General Data Protection Regulation (GDPR) \citep{GDPR2016}. Article 35 of the GDPR requires that for technologies that are \textit{``likely to result in a high risk to the rights and freedoms of natural persons,''} the controller shall conduct an assessment of the impact of the envisaged processing operations on the protection of personal data. Article 36 additionally requires ``prior consultation'' with the data protection supervisory authority for any data processing project for which a DPIA indicates a high risk with no identified set of measures to sufficiently mitigate the risk.

Hence, with the GDPR that came into force in 2016, a DPIA is now defined more formally, and data controllers are required to make an assessment of the impact on data protection, which means specifically in relation to the GDPR data protection principles stated in Article 5.
Formally, the right to data protection can be derived from Article 8 of the Charter of Fundamental Rights of the European Union (CFR) \citep{CFR2012}, whereas privacy is addressed in Article 7 of the CFR.
However, the differentiation between privacy and data protection may be regarded as not as important for DPIA and PIA since, in the European sense, PIA has never been strictly separated from data protection.
Moreover, Article 35 of the GDPR requires an assessment of technologies posing a high risk to the rights and freedoms of natural persons, which includes other fundamental rights (beyond the right to data protection) \citep{hallinan2020fundamentalrights}, including the right to privacy pursuant to Article 7 of the CFR.
Consequently, in the remainder of this paper, we do not strictly distinguish between PIAs and DPIAs, meaning that the investigation of PIAs in this paper will also address DPIAs.

The overall objective of this paper is to provide a Scoping Review (ScR) to analyse the state of PIAs in practice (``in the wild'') addressed by the research community and documented by the scientific literature. By PIA in practice, we mean empirical experience, application, and practical use of PIAs in real-world settings. We are particularly interested in elaborating on the positive aspects of PIAs (enablers, opportunities, drivers, and advantages) and the negative aspects (barriers, challenges, hindrances, and disadvantages), as reported by the scientific literature. 

Although prior work on PIAs sometimes regards them as commonplace \citep{edwards2012privacy} and widespread \citep{bayley2012privacy, stoddart2012auditing}, other studies have found that they are not commonly known and used by practitioners \citep{mckee2022pia, iwaya2023privacy}.
Additionally, empirical studies have also pointed to a low prevalence of PIAs for highly sensitive systems, such as mobile health apps \citep{iwaya2023mental}.
Such problems have motivated this study to survey the research field of PIAs in practice systematically, including the positive and negative aspects of PIAs that are experienced, and to gain more knowledge in order to understand the state of the art.

An earlier related investigation by \cite{wright2012state} on the state of the art of PIAs and their advantages, based on findings from the Privacy Impact Assessment Framework (PIAF) project, was (in contrast to our work) not published as a systematic review based on a well-defined protocol. It also dates back to 2012, before the GDPR was in force, and can thus not be regarded as up to date. Another descriptive field study on PIAs in practice is based on interviews with data protection officers and restricted to organisations in the Netherlands \citep{van2017privacy}. 
To the best of our knowledge, no related work on this topic has been published yet in the form of a systematic literature or scoping review that also considers recent research literature.

This ScR is based on 45 publications chosen following a formal review protocol and procedure.
As we show below in section \ref{sec:nature-of-evidence}, more than half of the studies on this topic (included in this review) were only published after 2016. However, publications before 2016 (before the year when the GDPR came into force) have rarely considered the DPIA context that the GDPR now gives. They were rather based on PIAs following their own ideas on what to assess, which were not necessarily the data protection principles of the GDPR, and not necessarily -- which is even more important -- risks related to the rights and freedoms of natural persons.

As a result, the contributions of this article are threefold: (i) a comprehensive synthesis of PIAs in practice, summarising existing research, main methodologies, and positive and negative aspects; (ii) a quality appraisal of existing qualitative and quantitative research on the topic; and (iii) a detailed discussion of identified research gaps and possible pathways for future research. These contributions benefit many stakeholders (e.g., privacy researchers, practitioners, law and policymakers) involved in the development and performance of PIAs. Privacy researchers can better understand the state of the art and pathways for future work on the practical aspects of PIAs. Practitioners responsible for carrying out PIAs in practice can further understand the barriers and enablers of the PIA process, select better approaches based on more knowledge and avoid pitfalls. Law and policymakers can also use the evidence-based insights of this review when developing guidelines, policies, and regulations, as well as when consulting with organisations.

The remainder of this article is structured as follows: Section \ref{sec:background} presents the background information and prior work; Section \ref{sec:methods} provides the details of the ScR methodology that guided this research; Section \ref{sec:results} presents the research results, including an account of the nature of existing evidence, a summary of the positive and negative aspects of PIAs in practice, and a critical appraisal of identified primary research; Section \ref{sec:threats-validity} describes the limitations of this ScR; Section \ref{sec:discussion} discusses the main findings and implications; and Section \ref{sec:conclusion} concludes the paper.

\section{Background}
\label{sec:background}
\subsection{PIAs are more than DPIAs}
The terms PIA and DPIA are often used interchangeably, but it is important to clarify that they differ. As briefly mentioned in the introduction, the concept of PIA far predates the EU framework GDPR and its legal requirement for DPIAs. PIAs come from a longer history, particularly in English-speaking countries, concerned with assessing the impact of technologies on people's privacy rights -- for a historical review of the origins of PIAs, we refer readers to the work of \citet{CLARKE2009123}. Nonetheless, GDPR has indeed motivated a great deal of research on PIA methodologies that assist practitioners in their attempts to systematise this process. Such methodologies are, however, explicitly named as ``PIAs,'' that is, adhering to the multi-dimensional concept of ``privacy'' rather than the narrower concept of ``data protection,'' as defined in the GDPR and Article 8 of the CFR. Some of the widely known PIA methodologies include:
\begin{itemize}
    \item New Zealander PIA Handbook \citep{opc2007piahandbook};
    \item German PIA guidelines \citep{bsi2011pia, bsi2011piarfid};
    \item French PIA Methodology \citep{cnil2015pia};
    \item Australian PIA Guide \citep{oaic2020piaguide}; and,
    \item ISO/IEC 29134 Guidelines for PIAs \citep{ISO29134}.
\end{itemize}

On the other hand, the DPIA is an instrument introduced by Article 35 of the GDPR \citep{GDPR2016}. Notice that the terminology starting with ``data protection'' was chosen consistently with the name of the law itself, that is, General ``Data Protection'' Regulation. However, well before the GDPR, policymakers and academics had already established the notion of PIAs. To some extent, DPIA becomes a new label for something already existing. In fact, when the Article 29 Working Party, the former body encompassing the data protection supervisory authorities from the European Union and predecessor of the European Data Protection Board, later created guidelines and recommendations on the conduction of ``DPIAs,'' they actually pointed to a series of well-known PIA methodologies (see Annex 1 of \citet{wp2017guidelines}). In this paper, given this context, we use the term PIAs more often, referring to a methodology that is common to many other countries and jurisdictions. The term DPIA is used to refer to PIAs as a legal requirement specific to Europe, based on the GDPR. Thus, DPIAs can be considered as a part of the broader area of PIA methodologies.

\subsection{Approaches to PIAs}
It is also important to clarify that PIA methodologies come in all shapes and sizes. For instance, some general methodologies were proposed by data protection authorities in different countries, such as previously mentioned ones in New Zealand \citep{opc2007piahandbook}, Germany \citep{bsi2011pia}, and France \citep{cnil2015pia}. Some methodologies can be considered industry- or application-specific, such as the PIA for RFID systems \citep{bsi2011piarfid}, the PIA for Smart Grids \citep{sgtf2018smartgrids}, or the PIA template for provenance systems \citep{reuben2016privacy}.

PIA methodologies can be differentiated regarding their comprehensiveness and depth of the assessment, sometimes deemed as \emph{lightweight} or \emph{full-fledged}. For instance, the PIA template for mobile health applications proposed by \citet{mantovani2017towards} consists of a concise questionnaire, which can be considered lightweight. The work of \citet{schneider2023persona} also proposed a software-based instrument that could serve as the low-threshold entry point for non-experts to start with DPIAs. However, while smaller businesses may be tempted to rely on such lightweight approaches, it should be noted that data protection supervisory authorities may not be satisfied with the result of such assessments. Therefore, full-fledged methodologies are more often recommended, which include many of the methodologies developed by privacy agencies (e.g., \citet{oaic2020piaguide, ico2018dpia, opc2007piahandbook, cnil2015pia}) and risk-based methodologies (e.g., \citet{cnil2015pia, oetzel2013systematic, bsi2011pia, bsi2011piarfid, ISO29134}) that borrow from traditional risk assessment theories, using the notions of threats and risks (i.e., with severity and likelihood levels).

This manuscript does not intend to review existing PIA methodologies comprehensively. Therefore, we refer the readers to important contributions in the area in which the authors have already theoretically compared and evaluated the most common PIA methodologies, as in the works of \citet{vemou2018evaluation, vemou2019evaluating} and \citet{bisztray2019privacy}. Furthermore, a systematic review of the scientific evidence on the \emph{evaluation} and \emph{validation} of PIA methodologies has also been presented by \citet{wairimu2024slr}. It is worth stressing, however, that the work of \cite{wairimu2024slr} dealt exclusively with \emph{scientific} PIA methodologies, which are seldom employed in practice, as also evidenced by the results of this ScR.

\section{Methods}
\label{sec:methods}
This study follows Scoping Review (ScR) protocols according to the Preferred Reporting Items for Systematic reviews and Meta-Analyses extension for Scoping Reviews (PRISMA-ScR) \citep{tricco2018prisma}, and adheres to the PRISMA-P \citep{moher2015preferred} checklist for the research protocol to ensure the completeness and transparency of the review process. An ScR is a type of systematic review used to map key concepts underpinning a research area and the main sources and types of evidence available \citep{arksey2005scoping}. This type of review emphasises a comprehensive coverage (i.e., breadth) of available literature \citep{arksey2005scoping} and is used to identify knowledge gaps, set research agendas, and identify implications for decision-making \citep{tricco2016scoping}. For this reason, it is worth mentioning that ScRs can also be considered as a preliminary type of systematic review that can later be narrowed down to more targeted topics for systematic literature reviews (SLRs) \citep{munn2018systematic}. The main phases and steps of this ScR are presented as follows.

\subsection{Phase I - Planning the ScR}

\subsubsection{Research Questions}
\label{sec:research-questions}
This study started with the objective of understanding the literature and research on the topic of PIAs in practice. Here, we are focusing on empirical evidence that deals with the experience, application, and practical use of PIAs in real-world settings. To meet this overarching objective, three guiding Research Questions (RQs) were formulated:
\begin{itemize}
    \item \textbf{RQ1:} \textit{What literature and research discuss and/or evaluate the use of PIAs in practice?} \textbf{Objective:} To identify the existing research and understand the types of studies and contributions in the literature, as well as the methodological approaches used, main venues, industry sectors, laws and regulations.
    \item \textbf{RQ2:} \textit{What are the main methodologies and positive and negative aspects of PIAs in practice as reported in the literature?} \textbf{Objective:} To identify relevant characteristics, such as the main methodologies reported in the studies, and compile shared experiences, focusing on understanding the main positive and negative aspects related to PIAs in practice.
    \item \textbf{RQ3:} \textit{What is the state of existing primary research studies on the topic?} \textbf{Objective:} To further examine and critically appraise existing primary research and to identify the studies (or the lack thereof) that have proposed valid and reliable instruments related to PIAs.
\end{itemize}

\subsubsection{Identifying the Need for the Review}
\label{sec:identify-scr-need}
Before undertaking the ScR, it is important to verify whether reviews have already been published on the topic. To do so, five scientific databases (Google Scholar, Scopus, IEEE Xplore, ACM Digital Library, and Web of Science) were searched using the keywords: \emph{``review''}, \emph{``survey''}, \emph{``privacy impact assessment*''}, and \emph{``data protection impact assessment*''}. No surveys or systematic reviews were found, confirming the need for this ScR and revealing a lack of secondary research on PIAs.

\subsubsection{Writing the ScR Protocol}
\label{sec:writing-protocol}
In preparation for the review, a ScR research protocol is written detailing all the systematic steps to be followed. This protocol was then discussed among the researchers and piloted, allowing us to refine search strategies and achieve a common understanding. The final protocol for this ScR is provided in \citep{iwaya2023protocol} together with a replication package \citep{iwaya2024repo}. Therefore, we refer the readers to these documents for extensive methodological details and supplementary materials.

\subsection{Phase II - Conducting the ScR}
\subsubsection{Information Sources and Search Process}
Due to their size and relevance to computer science and engineering, four databases were searched, including Scopus, Web of Science, IEEE Xplore, and ACM Digital Library.
The searches were done on 01/Nov/2023 without setting restrictions (e.g., year limits, publication types, etc.).
A structured search strategy was performed based on the RQs using the following general search string: \texttt{"privacy impact assessment*" OR "data protection impact assessment*"}.
Later, when reading the selected papers, backward and forward snowballing searches were also performed.
Although grey literature was also searched using the OpenGrey Database (\url{www.opengrey.eu}), no relevant studies were found.

\subsubsection{Eligibility Criteria}
\label{sec:eligibility-criteria}
To meet the eligibility criteria, publications must address both aspects: (1) PIAs or DPIAs; and their (2) practice, application, and empirical experience. Studies that discuss or more profoundly evaluate the use of PIAs in practice through the lens of many stakeholders (e.g., privacy officers, policymakers, data subjects, etc.) were also considered. However, the exclusion criteria refer to studies that only present a purely theoretical contribution, without any practical component (e.g., a study in a real-world setting or with real practitioners/stakeholders), such as studies that propose, compare, or validate PIAs solely on a hypothetical basis or ``illustrative'' use case scenarios. Furthermore, it is worth noting that Scoping Reviews should include all types of studies (e.g., opinion papers, experience papers, primary studies, evaluation studies), emphasising the breadth of literature and not being limited to exclusively primary research papers.

\subsubsection{Selection of Studies}
The results from each database were then exported separately (i.e., RIS files) and imported into the Rayyan software (\url{www.rayyan.ai}), a collaborative web tool that supports the screening process in systematic reviews. Rayyan automatically generates a list of duplicate entries, which the reviewers can easily check and manually remove. Two reviewers in double-blind mode performed this screening process using Rayyan. Before going through all the retrieved studies, the reviewers performed a calibration round to ensure a common understanding and minimise uncertainties about the eligibility criteria (see Section \ref{sec:eligibility-criteria}).
The selection of studies follows two main steps: Step 1) reading of titles and abstracts; and, Step 2) full reading of the studies. In Step 1, two reviewers used separate Rayyan accounts to analyse the whole list of retrieved studies. Based on the eligibility criteria, the reviewers decided whether to ``exclude'' or ``include'' the study for further analysis in Step 2. In case of disagreements, a third reviewer was called to establish the final decision. In Step 2, the full-texts of all studies included so far were downloaded and screened by the two reviewers again, following the eligibility criteria. If any reviewer deemed a study to be outside the scope, the study was discussed with a third reviewer for a final decision.

\subsubsection{Data Extraction and Data Charting Processes}
A data extraction form was created, tested, and discussed among the research team -- available in \citet{iwaya2024repo}. This form facilitates the consistent extraction of relevant data from each publication in separate files. This data extraction step later facilitated the data charting process. Data extraction was an iterative process, allowing reviewers to critique, agree, and update the data extraction form as needed. Only direct quotes were extracted from the studies to ensure accuracy. Established classification schemes were used to classify (a) Types of Research and (b) Types of Contributions. The categories proposed by \citet{wieringa2006requirements} and  \citet{creswell2017research} were used for the research types. Furthermore, the classification proposed by \citet{shaw2003writing} is used for contribution types. Nonetheless, notes were always included in the form to justify any classification choices.

\subsection{Phase III - Reporting the ScR}
\subsubsection{Synthesis of Results}
The synthesis of results aims to combine data from included studies into a coherent narrative using tables, charts, and classifications. This is facilitated by the data charting performed parallel to the data extraction steps. Two groups of data were generated: quantitative data (e.g., year of publication, publications per author, citations) and qualitative data (e.g., type of publication, type of research, type of contribution). Reviewers could then perform the synthesis via spreadsheet and text software (LibreOffice Calc and Writer), generating tables and figures, interpreting findings, and providing a descriptive mapping of the body of knowledge. Thus, a coherent narrative was written to convey all results, interpret the main findings, and discuss research gaps.

As part of the narrative summary, we derived from the literature a series of positive and negative aspects of PIAs in practice to answer RQ2. To do so, excerpts and quotes from the studies that reflected on such positive and negative aspects were captured in the data extraction step and later analysed by the researchers. A thematic analysis methodology \citep{braun2006thematic} was employed to code all the extracted materials qualitatively -- the NVivo software was used. This process was led by one researcher, initiating it with open coding under two broad themes (i.e., (1) positive and (2) negative aspects). With the help of two other researchers, all the generated codes were then revised for a shared understanding of their names, descriptions, and consistency with the original excerpts and quotes. The codes were then organised into sub-themes. The group of three researchers iterated continuously through the list of sub-themes and codes until a consensus was reached on the resulting sub-themes.

\subsubsection{Critical Appraisal of Primary Research}
During the synthesis, it was found that a critical appraisal of primary research was justified. To assess the methodological quality of the studies, two critical appraisal tools from the Center for Evidence-Based Management (CEBMa) were used: 1) \textit{Checklist for Qualitative Studies} \citep{cebma2014cap-quali}; and 2) \textit{Checklist for Cross-Sectional Studies (Surveys)} \citep{cebma2014cap-surveys}. The appraisals were done by one reviewer and checked by the group.

\subsubsection{Reporting Results}
This paper reports the results of the ScR, presenting them in line with the RQs (in Section \ref{sec:research-questions}).
The \textit{Synthesis of the Results} is presented in Sections \ref{sec:nature-of-evidence} and \ref{sec:scientific-reports}, including findings on the nature of existing research and the positive and negative aspects of PIAs in practice.
The \textit{Critical Appraisal of Primary Research} is provided in Section \ref{sec:state-primary-research}, detailing the quality of primary research studies.

\subsection{Results of the Selection Process}
\begin{figure}[htbp]
  \centering
  \includegraphics[width=0.85\linewidth]{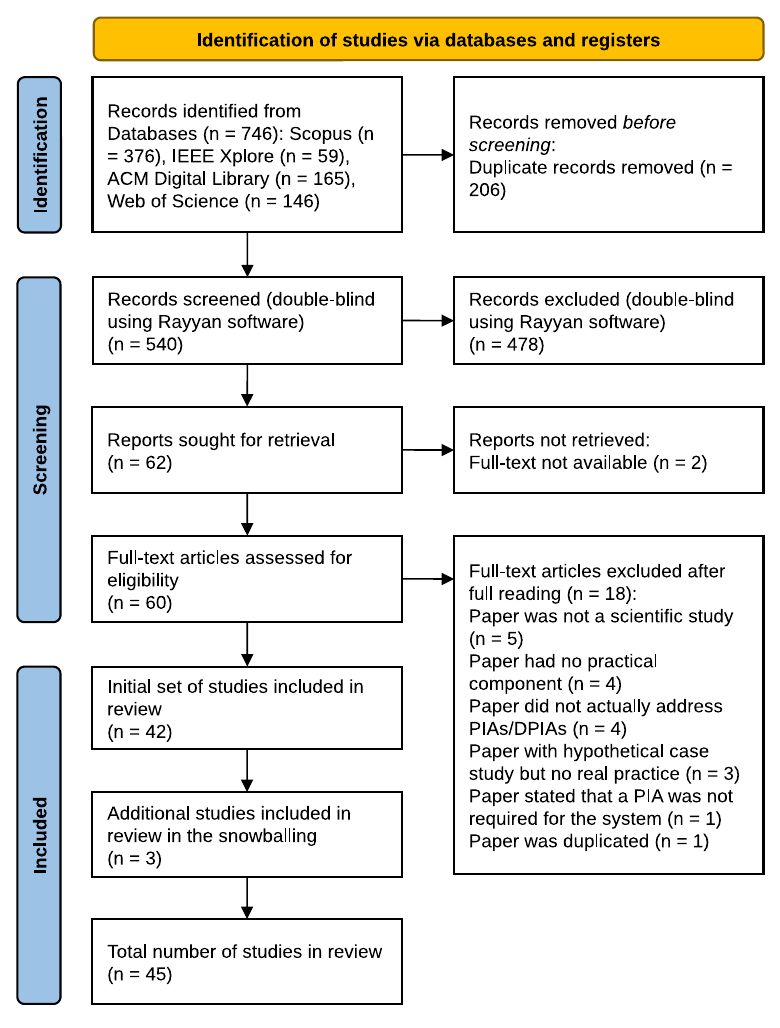}
  \caption{PRISMA flowchart for this ScR.}
  \label{fig:prisma-flow}
\end{figure}

Figure \ref{fig:prisma-flow} provides an overview of the selection process for this scoping review.
An initial total of 746 studies were identified through the search process.
The screening process using Rayyan was performed by two reviewers in double-blind mode, resulting in the identification of 60 relevant studies.
Only two studies were not accessible through the researchers' institutional libraries.
In such cases, we contacted the authors to request the publication but received no responses.
After the screening process, 42 papers were initially included in this review.
Using Google Scholar, we checked for all the papers that cited the selected studies (i.e., forward snowballing).
This process revealed another two studies that the authors deemed relevant.
While reading the full texts, we also looked for studies mentioned or referenced in each publication (i.e., backward snowballing).
Although four studies were identified through backward snowballing, two of them were not found in any databases (not even the bibliographic entries), and one of them was not accessible, so the authors were contacted, but they were also unable to locate the full texts.
One additional study was included as part of the backward snowballing process.
Finally, 45 studies were included in the ScR (complete list in \ref{app:list-of-papers}).

\section{Results}
\label{sec:results}
In this section, the results are organised according to each of our RQs. First, RQ1 is addressed in Section \ref{sec:nature-of-evidence} by presenting the nature of the literature and evidence of the topic, describing several demographics of the selected studies. The results related to RQ2 are presented in Section \ref{sec:scientific-reports}, describing the main PIAs methodologies and a thematic framework of the positive and negative aspects in practice. Lastly, concerning RQ3, the identified primary research studies are critically appraised in Section \ref{sec:state-primary-research}.

\subsection{The Nature of Evidence on the Topic}
\label{sec:nature-of-evidence}
\subsubsection{Frequency, Types of Research, and Contributions}
\begin{figure}[ht]
     \centering
     \begin{subfigure}[b]{0.5\textwidth}
         \centering
         \includegraphics[width=\textwidth]{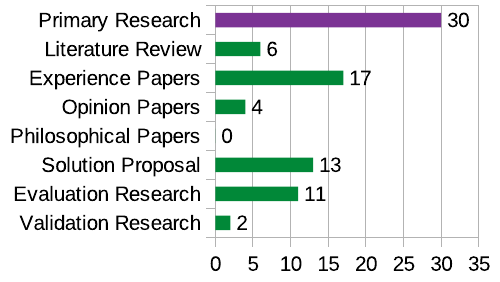}
         \caption{Types of Research}
         \label{fig:research-types}
     \end{subfigure}
     \hfill
     \begin{subfigure}[b]{0.4\textwidth}
         \centering
         \includegraphics[width=\textwidth]{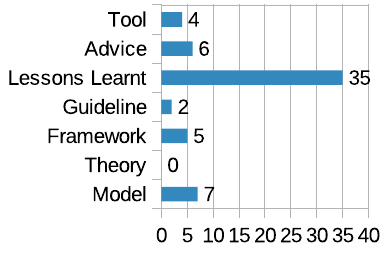}
         \caption{Types of Contributions}
         \label{fig:contribution-types}
     \end{subfigure}
    \caption{Chart (a) number of studies categorised by Research Types. Chart (b) number of studies for different types of contributions.}
    \label{fig:res-contrb-types}
\end{figure}

The following subsections aim to answer RQ1 by describing the key characteristics of the studies included in the ScR.
The studies were categorised using the classification schemes for research types proposed by \citet{wieringa2006requirements} and \citet{creswell2017research}.
As shown in Figure \ref{fig:research-types}, Primary Research is the main type of research with 66.7\% of the studies of a qualitative or quantitative nature -- primary research studies are further analysed in Section \ref{sec:state-primary-research}. The second most common type of research is Experience Papers (37.8\%). Such studies describe an author's first-hand experience with dealing with PIAs in specific jurisdictions (e.g., Australia \citep{clarke2016pia} and the UK \citep{, warren2012privacy}), private organisations (e.g., Nokia \citep{brautigam2012pia}, Siemens \citep{thoma2012siemens}, and Vodafone \citep{deadman2012vodafone}), or for specific projects (e.g., a data sharing platform \citep{horak2019gdpr}, the BIRO project \citep{diiorio2009pia}). Many studies (28.9\%) also propose new solutions, such as suggesting new or enhanced PIA frameworks and processes \citep{ahmadian2018supporting, todde2020methodology, henriksen-bulmer2020dpia, kroener2021agile, friedewald2022dpias, mckee2022pia, schneider2023persona}. Evaluation Research remains incipient, with only a few studies proposing new methodologies performing their practical evaluations (i.e., \citet{ahmadian2018supporting, todde2020methodology, henriksen-bulmer2020dpia, friedewald2022dpias}), while other studies evaluate public PIA reports and documents \citep{clarke2016pia, wadhwa2012privacy, shin2017analysis}, or the use a specific PIA methodology (i.e., CNIL PIA tool in \citet{campanile2022evaluating, alaqra2021machine, alaqra2023transparency}).

\begin{figure}[ht]
    \centering
    \includegraphics[width=0.6\textwidth]{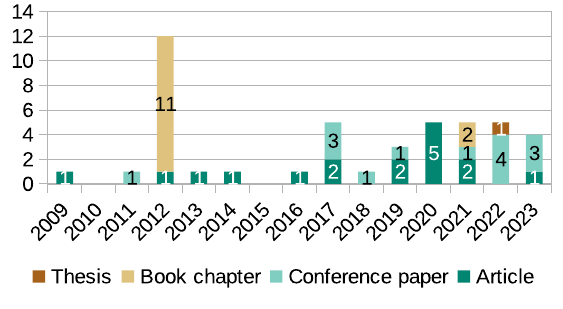}
    \caption{Publications published per year.}
    \label{fig:pubs-year}
\end{figure}

It is important to clarify that six studies were also classified as Literature Reviews since their authors explicitly mentioned it as part of the objectives (i.e., \citet{diiorio2009pia, vandercruysse2020typology, sharma2017strategy, wright2012findings, mckee2022pia, schneider2023persona}). However, these studies have neither performed systematic reviews as part of their research nor addressed the topic of PIAs ``in the wild.'' Another clarification refers to the studies classified as Opinion Papers. These studies present personal opinions of PIAs in practice primarily based on first-hand experience and without relying on related work or research methodologies (i.e., \citet{bamberger2012pia, bayley2012privacy, stewart2012privacy, rehak2022analysis}). Although some types of systematic review methodologies would exclude such types of publications, ScRs purposefully include such studies since they seek to map a research area comprehensively, as discussed in \citet{munn2018systematic}.

For the contribution types, studies were categorised using the classification proposed by \citet{shaw2003writing}.
As shown in Figure \ref{fig:contribution-types}, most studies (77.8\%) contributed with Lessons Learnt drawn from their results.
Five studies (11.1\%) also contributed with Frameworks, that is, PIA frameworks or methodologies \citep{ahmadian2018supporting, todde2020methodology, henriksen-bulmer2020dpia, kroener2021agile, friedewald2022dpias}, and seven studies (15.6\%) with various kinds of Models (e.g., PIA process model \citep{mckee2022pia}, system model \citep{diiorio2009pia}, threat and control models \citep{iwaya2019mobile, zamorano2022privacy}, or a typology \citep{vandercruysse2020typology}).

\subsubsection{Countries and Publication Venues}
By reviewing the affiliations of the authors of the included studies, we found that 12 studies (26.7\%) came from the United Kingdom, followed by Germany with five studies (11.1\%), the Netherlands and Italy with four studies each (8.9\%), and the United States, Belgium, Australia and France with three studies each (6.7\%). Institutions of 14 other countries were accounted for in the studies but with only two or fewer publications. Besides, the publications come from a heterogeneous number of venues, with 32 studies published in various conference proceedings, journals, or as doctoral theses. Eleven studies are chapters of the book \citet{wright2012privacy}, and two studies come from the journal of \textit{International Data Privacy Law}.

\subsubsection{Industry Sectors, Laws and Regulations}
\begin{figure}[ht]
    \centering
    \includegraphics[width=0.5\textwidth]{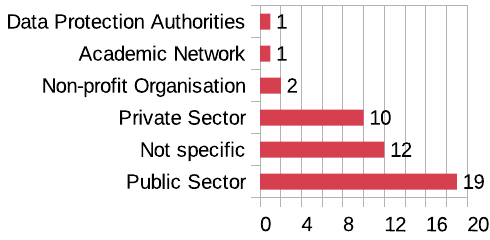}
    \caption{Types of Organisations.}
    \label{fig:organisations}
\end{figure}

During the data extraction stage, relevant information concerning the application context of the studies was captured, generating a classification for types of organisations, as shown in Figure \ref{fig:organisations}.
Most studies (42.2\%) were applied to Public Sector organisations, such as studies focused on PIA processes in government agencies \citep{bamberger2012pia, stoddart2012auditing, shin2017analysis}, public health and healthcare institutions \citep{rehak2022analysis, rehak2022processing, todde2020methodology}, or national security and police agencies \citep{clarke2016pia, rajamaki2021design, basseyyar2020pia, zamorano2022privacy}.
Many studies (26.7\%) were not focused on a specific context, addressing the topic of PIAs more broadly and involving various stakeholders and practitioners, such as \citet{ferra2020challenges, vandercruysse2020typology, vandercruysse2021dpia, wright2014integrating, schneider2023persona}.
The private sector was represented in 10 studies (22.2\%), such as in reports from Nokia \citep{brautigam2012pia}, Vodafone \citep{deadman2012vodafone}, Siemens \citep{thoma2012siemens}, among other sectors, for instance banking \citep{sharma2017strategy}, medical device manufacturers \citep{mckee2022pia}, telecommunications \citep{alaqra2023transparency}, and mobile app companies \citep{iwaya2023mental}.

\begin{figure}[htbp]
    \begin{subfigure}[b]{0.5\textwidth}
         \centering
         \includegraphics[width=\textwidth]{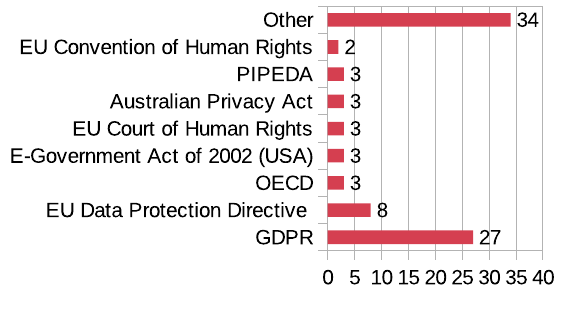}
         \caption{Privacy Regulations}
         \label{fig:regulations}
    \end{subfigure}
    \hfill
    \begin{subfigure}[b]{0.4\textwidth}
         \centering
         \includegraphics[width=\textwidth]{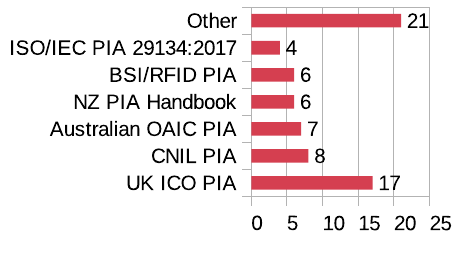}
         \caption{PIA Frameworks}
         \label{fig:pia-methods}
    \end{subfigure}
    \caption{Chart (a) most mentioned privacy regulations, and chart (b) most mentioned PIA/DPIA frameworks and methodologies.}
    \label{fig:methods-regulations-organisations}
\end{figure}

Regarding privacy laws and regulations, the studies collectively mention many legal references, but the most commonly mentioned is the EU GDPR (60\%).
Other privacy regulations were mentioned to a lesser extent, such as the EU Data Protection Directive (17.8\%) that preceded the EU GDPR, and the OECD privacy principles (6.7\%), among others.

\subsection{The Scientific Reports on PIAs in Practice}
\label{sec:scientific-reports}

\subsubsection{PIA Frameworks}
Many PIA frameworks (i.e., full-fledged methodologies or guidelines) that are mentioned or used in the studies were also identified.
As shown in Figure \ref{fig:pia-methods}, the most popular PIA framework was the one from the UK Information Commissioner's Office, which was mentioned in 17 studies (37.8\%).
Other well-known PIA frameworks were also mentioned: the French PIA Methodology \citep{cnil2015pia} (17.8\%) from the Commission nationale de l'informatique et des libertés; the Australian PIA Guide \citep{oaic2020piaguide} (15.6\%) from the Office of Australian Information Commissioner; the New Zealander PIA Handbook \citep{opc2007piahandbook} (13.3\%) from the Office of the Privacy Commissioner; the German PIA guidelines \citep{bsi2011pia, bsi2011piarfid} (13.3\%) provided by the Bundesamt f\"{u}r Sicherheit in der Informationstechnik and based on the work of \citet{oetzel2013systematic}; and the ISO/IEC 29134 Guidelines for PIAs \citep{ISO29134} (8.9\%).
Several PIA frameworks were categorised as Other (46.7\%) since they were mentioned in only three or fewer studies.

Among the selected studies in this ScR, eight publications introduced a novel PIA or DPIA methodology/framework as part of their research (see Table \ref{tab:new-pia-methods}).
In 2013, the work of \citet{wright2013introducing} introduced the Privacy Impact Assessment Framework (PIAF).
The other seven studies, published between 2018 and 2023 (after GDPR came into effect), proposed other solutions, such as the PIA by model-based privacy analysis \citep{ahmadian2018supporting} and the persona-oriented DPIA \citep{schneider2023persona}.
Also, eight studies describe their use and experience with existing PIA frameworks, as shown in Table \ref{tab:use-existing-pias}.
The most used methodology among these studies is the CNIL PIA tool, which was also used as a basis for the PAPAYA PIA Tool discussed in \citet{alaqra2023transparency} and \citet{pulls2019risk}.

\begin{table}
\centering
\footnotesize
\caption{Studies that introduced new PIA/DPIA methodologies or frameworks.}
\label{tab:new-pia-methods}
    \begin{tabular}{|l|l|}
         \hline
         \textbf{New methodology} & \textbf{Ref.} \\
         \hline \hline
         PIA by Model-Based Privacy Analysis & \citet{ahmadian2018supporting} \\
         Privacy Impact Assessment Framework & \citet{wright2013introducing} \\
         DPIA methodology for healthcare systems & \citet{todde2020methodology} \\
         DPIA Data Wheel methodology & \citet{henriksen-bulmer2020dpia} \\
         Agile E/PIA methodology & \citet{kroener2021agile} \\
         DPIA framework & \citet{friedewald2022dpias} \\
         Privacy Assessment Methodology Model & \citet{mckee2022pia} \\
         Persona-Oriented DPIA for SMEs & \citet{schneider2023persona} \\
         \hline
    \end{tabular}
\end{table}

\begin{table}
\centering
\footnotesize
\caption{Studies that adopted an existing PIA methodology or framework.}
\label{tab:use-existing-pias}
    \begin{tabular}{|l|p{0.7\textwidth}|}
         \hline
         \textbf{Methodology} & \textbf{Ref.} \\
         \hline \hline
         CNIL PIA & \citet{campanile2022evaluating, rajamaki2021design, dashti2021can, alaqra2021machine} \\
         PAPAYA PIA Tool & \citet{alaqra2023transparency} \\
         BSI/RFID PIA & \citet{iwaya2019mobile} \\
         UK ICO PIA & \citet{wright2014integrating} \\
         AU OAIC PIA & \citet{pribadi2017regulatory} \\
         \hline
    \end{tabular}
\end{table}

Apart from that, five studies have described \textit{ad-hoc} or ``in-house'' PIA methodologies that were used in their organisations.
For instance, the works of \citet{horak2019gdpr}, \citet{diiorio2009pia}, and \citet{basseyyar2020pia} describe the conduction and experience with PIAs that were devised based on existing regulations and applied \textit{ad-hoc} in their specific projects.
Other industries describe their in-house PIA approaches, such as Nokia's privacy impact and security assessment (PISA) template \citep{brautigam2012pia} and Vodafone's Strategic PIA as part of their privacy risk management system.

\subsubsection{On the Positive and Negative Aspects of PIAs in Practice}
\label{sec:positives-and-negatives}
\begin{figure}[ht]
  \centering
  \includegraphics[width=\linewidth]{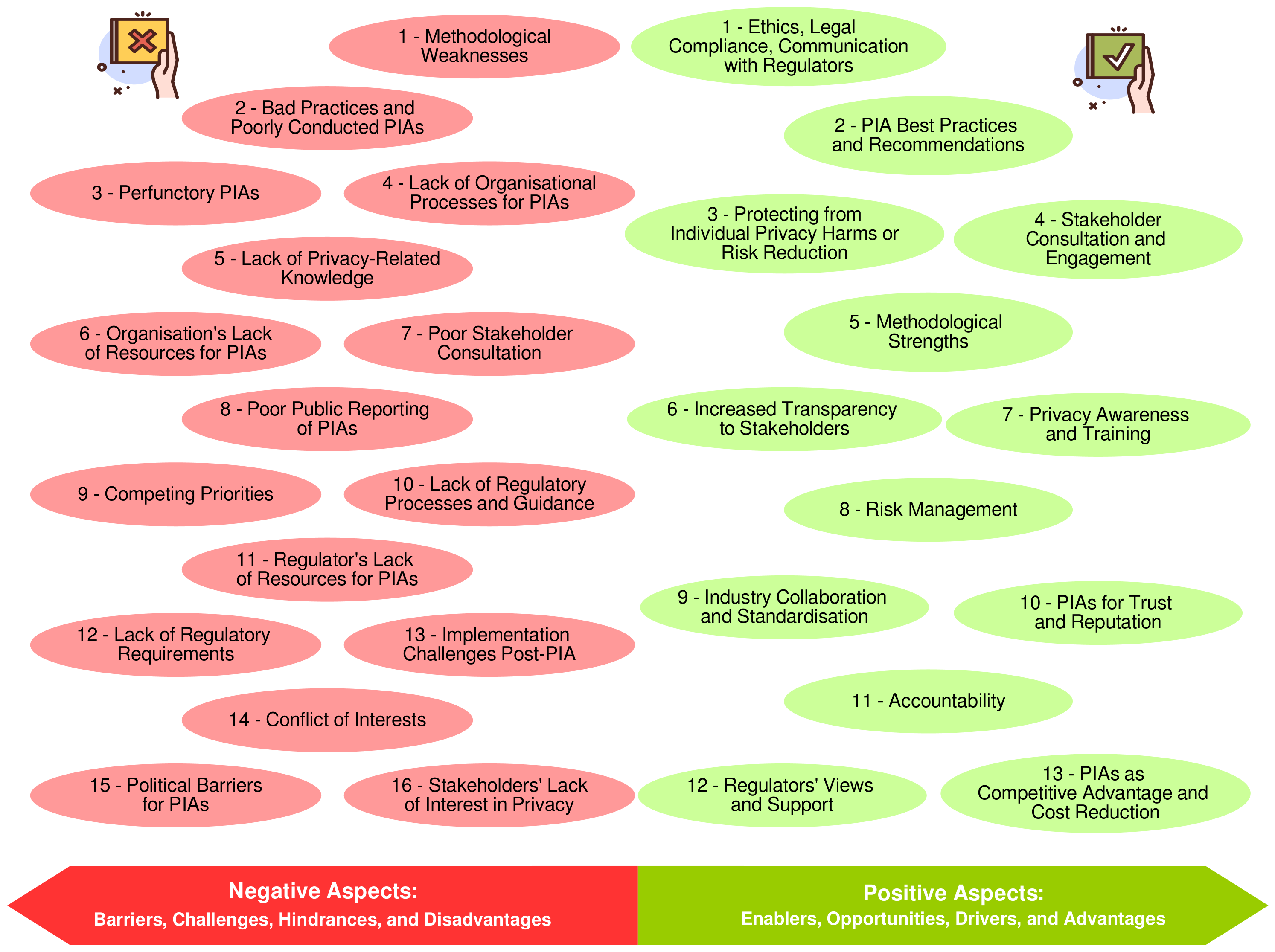}
  \caption{An overview of the sub-themes on the Positive and Negative Aspects related to PIAs in practice as reported in the studies.}
  \label{fig:positives-and-negatives}
\end{figure}

This subsection focuses on answering RQ2 concerning the positive and negative aspects of PIAs in practice, as reported in the ScR studies.
Figure \ref{fig:positives-and-negatives} shows an overall picture of the main aspects identified by the authors from reviewing the studies.
Even though such aspects come from the authors' efforts to synthesise the content, it is worth stressing that this list is not meant to be an exhaustive enumeration of aspects but rather to be seen as a high-level overview.
The aspects are numbered in order of prevalence, that is, the aspect is covered in more studies; however, this should not be interpreted as a prioritisation or ranking of importance.
As part of the replication package \citep{iwaya2024repo}, a summary of the generated codebook is provided as supplementary material, containing exemplary quotes and references that helped us form each sub-theme.
A compilation of positive (P) and negative (N) aspects is presented in what follows.

\subsubsection*{P1 - Ethics, Legal Compliance, Communication with Regulators}
\label{sec:p1}
\begin{landscape}
\begin{figure}[ht]
  \centering
  \includegraphics[width=\linewidth]{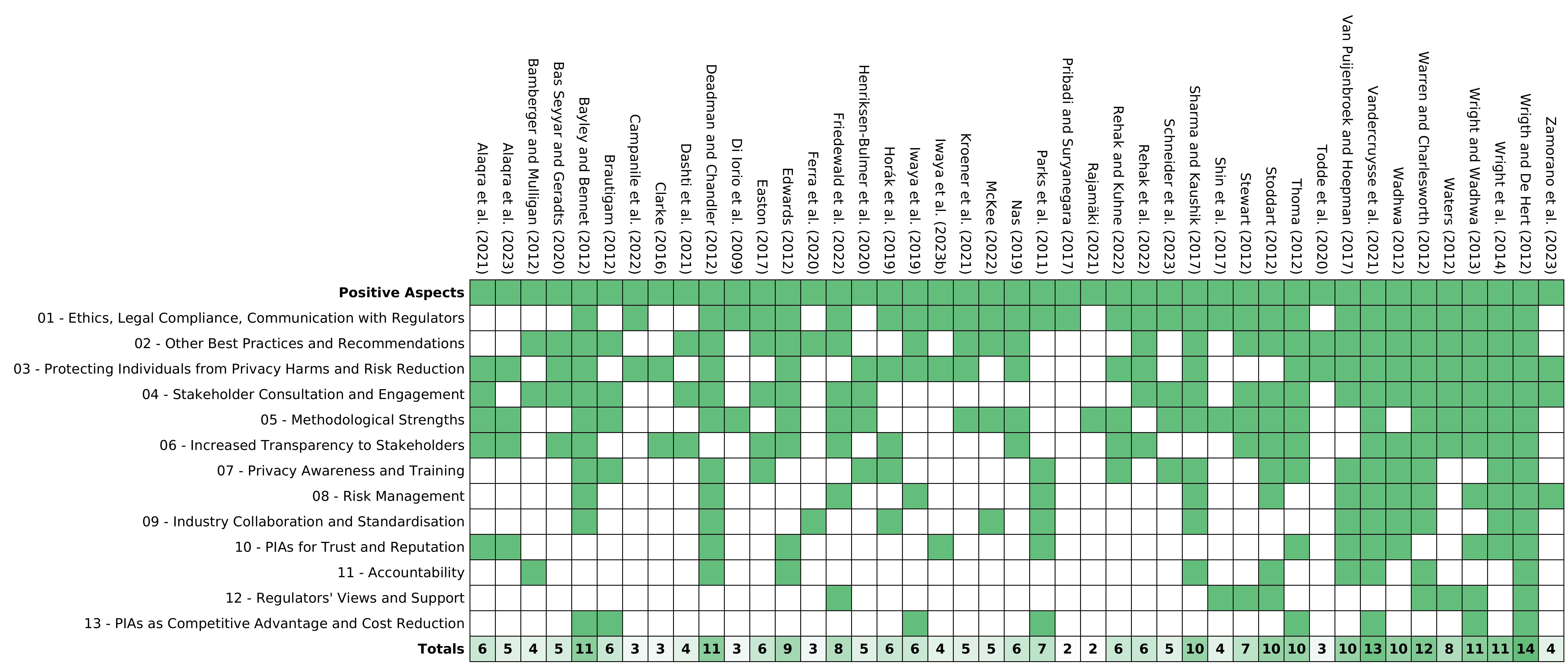}
  \caption{Overview of studies distributed across the sub-themes of Positive Aspects related to PIAs in practice.}
  \label{fig:positive-distribution}
\end{figure}
\end{landscape}

Starting with the positive aspects, the identified sub-themes are distributed over a total of 43 studies, as shown in Figure \ref{fig:positive-distribution}.
Most studies stressed the benefits of \textbf{PIAs for meeting regulatory compliance} when developing or deploying systems.
PIAs are regarded as established risk analysis methodologies that enable organisations to pursue their commercial strategies whilst satisfying people's privacy expectations and \textbf{meeting legal and ethical obligations} \citep{deadman2012vodafone}.
In addition, it was also often mentioned that PIAs can be leveraged to better communicate with regulators (e.g., DPAs or privacy commissioners).
PIAs can act as \textbf{accountability mechanisms in the review or audit of implemented mitigation measures} \citep{bayley2012privacy}, facilitating the investigations for regulators.
At the same time, organisations can use the PIAs and documented processes for \textbf{demonstrating compliance with regulators}, data subjects, and platform users \citep{horak2019gdpr}.
Ultimately, proper PIAs can be used as \textbf{a basis for discussion with regulators} as well as other stakeholders (e.g., researchers, journalists, civil societies, the general public) \citep{rehak2022processing}.

\subsubsection*{P2 - PIA Best Practices and Recommendations}
\label{sec:p2}
Several best practices and recommendations for high-quality PIAs were reflected in the studies.
Among them, it is essential to understand \textbf{PIAs as processes and living documents} that should be iteratively refined whenever privacy objectives shift in the system design \citep{basseyyar2020pia}, and periodically reviewed for changes of assumptions or when new privacy threats are discovered \citep{iwaya2019mobile}.
PIAs are only effective if their outcomes can be realised, so it is also important to \textbf{start early to maximise results} when project changes are still feasible.
At an early stage, the project team can consider warnings in the PIA, allowing them to rethink the design, focusing on purposeful features that are less privacy-invasive \citep{waters2012privacy}.
Similarly, if an organisation is acquiring new information systems, PIAs can better inform the project team if performed before the deployment process in the real environment \citep{todde2020methodology}.

Practitioners also argued that in order to be effective, PIAs need to \textbf{define the scope of analysis and limitations} with proportional depth to many crucial variables, such as the size of the organisation, sensitivity of the data and processing operations, forms of risks and intrusiveness of the technology \citep{warren2012privacy}.
The scope constraints of the project must be clearly described, and the PIA report should only account for such precise descriptions, protecting assessors from potential scope creep in the future \citep{edwards2012privacy}.

Another strong recommendation by practitioners is to ensure the \textbf{independence of privacy experts and assessors}, whether they are internal (e.g., privacy officers, experts, hired consultants) or external (e.g., regulators) to the organisation.
Regulators clearly have an obligation to be independent of the organisations since they represent the public interest and are expected to protect the community \citep{stewart2012privacy}.
More importantly, the internal representatives in the organisation responsible for the assessment should also have enough independence so that when highlighting challenges, legal limitations, and risks, their assessment cannot be simply overturned or dismissed by other members \citep{thoma2012siemens}.
Practitioners also emphasised the importance of \textbf{developing a strong process for PIAs}, with proper scoping and project management methods for handling the assessments in a sensible and efficient manner \citep{vandercruysse2021dpia}.

Finally, practitioners indicated that they can \textbf{reuse and share past PIAs, tools, and catalogues} to help others inside or outside the organisation.
Those conducting large numbers of PIAs can compile supporting materials of common privacy risks and mitigation strategies, improving the efficiency of future PIAs and accumulating knowledge within the organisation \citep{bayley2012privacy}.
Nonetheless, it is also possible to take advantage of well-maintained catalogues of privacy threats and controls that supervisory authorities have already created \citep{friedewald2022dpias}, for instance, CNIL's PIA templates \citep{cnil2018templates} and knowledge bases \citep{cnil2018knowledge}.

\subsubsection*{P3 - Protecting Individuals from Privacy Harms and Risk Reduction}
\label{sec:p3}
Several studies described using \textbf{PIAs for better privacy, minimising privacy harms by identifying threats and appropriate controls} with the ultimate focus on the data subjects.
PIAs helped in visualising impact levels (e.g., matrices for risk likelihood and severity), raising awareness among developers and finding better ways to handle personal data \citep{campanile2022evaluating}.
They also compel project members to justify the data collection and processing operations, balancing risks, costs for implementing controls, and resulting benefits \citep{kroener2021agile}.
In doing so, PIAs bridge a gap between legal principles and engineering practices by providing a systematic way of identifying risks and controls, giving developers a clearer path toward solving privacy issues and protecting individuals \citep{iwaya2019mobile}.

\subsubsection*{P4 - Stakeholder Consultation and Engagement}
\label{sec:p4}
High-quality PIAs require good consultation with both \textbf{internal} and \textbf{external} stakeholders.
Among the internal stakeholders, \textbf{senior and privacy experts} with diverse disciplinary skills can be involved, or preferably, embedded within operational units of the organisation, taking advantage of the interdisciplinary nature of PIA tools for generating ``bottom-up'' decision-making \citep{bamberger2012pia}.
PIA practitioners can also leverage internal \textbf{collaboration cross-teams and with domain experts}, learning from experience across departmental boundaries and getting directly involved with developers and main internal users.
The collaboration between developers, domain experts, and PIA practitioners is crucial in the threat elicitation process and helps raise privacy awareness in the teams \citep{basseyyar2020pia}.
Another important internal aspect is getting \textbf{management support} throughout the PIA process, that is, ``buy-in'' from senior management, also developing a culture of privacy within the organisation \citep{wright2014integrating}.
Similarly, \textbf{engaging project members} was also regarded as crucial, that is, ensuring that people involved do not see the PIA as a constraining exercise or a legal obstacle and, instead, focus on the wider PIA benefits such as transparency, confidence, and process streamlining that can directly benefit the project, the system under development, or the organisation \citep{easton2017analysing}.

A significant part of the PIAs also involves \textbf{engaging with external stakeholders} that can help to inform, seek agreement, and provide feedback on the policies of the organisation \citep{deadman2012vodafone}.
More specifically, it is of utmost importance to leverage \textbf{public consultations and involve data subjects} (or their representatives) as a thorough basis for assessing privacy risks, seeking heterogeneous views \citep{friedewald2022dpias}.
Studies also recognised the expertise of the regulators and the usefulness of \textbf{external PIA reviews} and feedback, in which organisations can seek comments to further understand and manage privacy risks to deliver good privacy outcomes to individuals \citep{stewart2012privacy}.

\subsubsection*{P5 - Methodological Strengths}
\label{sec:p5}
PIA frameworks are often acclaimed for their methodological strengths.
Some consider \textbf{PIAs as a privacy-by-design approach} since the methodologies offer a general solution to build robust privacy-protective information systems \citep{diiorio2009pia}.
The \textbf{continuous improvement and streamlining of PIA frameworks}, methodologies, and software tools have also made PIA more practical and improved collaboration among practitioners and other stakeholders \citep{rajamaki2021design}.
Security researchers also refer to \textbf{PIAs as systematic and sound methodologies for risk analysis} that can be used for rigorous evaluation of systems, and final reports can be disseminated for public scrutiny \citep{rehak2022processing}.
Throughout the years, \textbf{PIAs have been considered a widespread practice} \citep{bayley2012privacy, stoddart2012auditing}, being commissioned and prepared in a wide range of initiatives and innovation proposal \citep{edwards2012privacy}, which has also helped to establish the methodology among practitioners and policymakers.

\subsubsection*{P6 - Increased Transparency to Stakeholders}
\label{sec:p6}
Using \textbf{PIAs for increased transparency} was also reported as a key advantage. As noted by \citet{clarke2016pia}, the conduct of proper PIAs has the benefit of forcing a degree of transparency, enabling informed decision-making by stakeholders that can more easily identify privacy issues and impose appropriate controls.
To achieve higher levels of transparency, \textbf{publishing the public PIA reports} of systems has also been suggested as a way to deal with privacy concerns through public debate.
In this way, it is also possible to use \textbf{PIAs as a basis for societal discussion of privacy risks}, enabling the informing and involvement of the media, political parties, and the broad population \citep{nas2019data}.
Nonetheless, such public PIA reports should be made easily accessible instead of having them buried on a website. An alternative is to \textbf{leverage central PIA repositories} that could be managed by a supervisory authority \citep{wright2012findings}.

\subsubsection*{P7 - Privacy Awareness and Training}
\label{sec:p7}
It has been reported that companies can use the \textbf{PIA self-assessment for increased privacy awareness in the organisation}, especially when using contextual questions as part of the PIAs, clarifying the meanings of privacy risks and its nuances in a specific context \citep{henriksen-bulmer2020dpia}.
Some studies also discussed the link between \textbf{PIAs and the organisational privacy culture}, emphasising the benefits of proactive approaches to ensure privacy, such as the annual refresh of PIAs, trigger-based updates, and promoting a privacy culture \citep{sharma2017strategy}.
In addition, organisations should \textbf{ensure that staff has PIA and privacy training}.
Specialised PIA training and basic privacy training can help bridge knowledge gaps in the organisation by equipping more professionals to participate in PIAs instead of stretching the resources of a few qualified staff \citep{bayley2012privacy}.

\subsubsection*{P8 - Risk Management}
\label{sec:p8}
Many studies have highlighted the need to integrate \textbf{PIAs as part of the risk management process} \citep{wright2014integrating}.
The organisational risk management approach can readily use PIA frameworks as effective tools for assessing different types of privacy risks \citep{parks2011understanding}.
Furthermore, the use of \textbf{PIAs helps to identify the people responsible for treating issues and risks in the organisation}. Whilst seeking agreement and prioritising actions, the PIA process also brings up the discussion on the correct people within business units that are responsible and can effectively treat privacy risks (i.e., the privacy risk owners), which are likely to be different from privacy officers of managers \citep{deadman2012vodafone}.

\subsubsection*{P9 - Industry Collaboration and Standardisation}
\label{sec:p9}
Industry practitioners \textbf{sharing PIA experience and tools with others outside the organisation} were found to be a positive aspect for increasing knowledge levels and confidence in PIA practices \citep{ferra2020challenges}.
Studies have shown that many \textbf{industry professionals see benefits and plan to use PIAs} once they receive basic training on the methodologies \citep{mckee2022pia}.
Standardisation initiatives have also helped to establish PIA frameworks in the industry further, for instance, ISO/IEC 29134:2023 \citep{ISO29134}.

\subsubsection*{P10 - PIAs for Trust and Reputation}
\label{sec:p10}
Organisations can also leverage \textbf{PIAs for building trust and reputation}, using them as a foundation of their broader privacy strategies \citep{deadman2012vodafone}.
A strong reputation for privacy helps companies garner citizens' trust, especially when they publish PIA reports, showing more transparency and allowing business associates to judge procedures on a more technical level \citep{vandercruysse2021dpia}.

\subsubsection*{P11 - Accountability}
\label{sec:p11}
When adopting PIA practices, it is critical that organisations \textbf{develop mechanisms for accountability, monitoring, and internal or external oversight} to see benefits to their full potential, as advised by scholars and policymakers \citep{bamberger2012pia}.
For instance, a mechanism or strategy includes the signing off of the PIA report by critical parties, such as, project owners, line managers, privacy and security officers, or executives \citep{stoddart2012auditing, sharma2017strategy, van2017privacy}. This approach improved the involvement of the key parties and the quality of the PIA reports \citep{van2017privacy}.

\subsubsection*{P12 - Regulators’ Views and Support}
\label{sec:p12}
Some studies reported that support and \textbf{clearer guidelines for PIAs from data protection supervisory authorities are helpful}, allowing practitioners to have more precise and clear explanations \citep{friedewald2022dpias}. An example is the guideline that further explains the notion of data processing operations with ``high risk'' to the rights and freedoms of individuals \citep{wp2017guidelines}, therefore providing clearer criteria for assessors.
Studies focused on the Data Protection Authorities (DPAs) also indicate that the \textbf{regulators play a strong advocacy role} in ensuring that PIAs are properly done, that they are independently reviewed, and that safeguards are implemented, thus impacting the citizens' and consumers' trust in organisations \citep{wright2013introducing}.

\subsubsection*{P13 - PIAs as Competitive Advantage and Cost Reduction}
\label{sec:p13}
Even though privacy requires significant investments from organisations, experts mostly agree that \textbf{the benefits of a PIA far outweigh the costs} \citep{brautigam2012pia, wright2013introducing}.
As a privacy-by-design strategy, PIAs allow issues to be identified at the conception stage when design changes are significantly simpler and less costly if compared to later stages of software development (such as testing and deployment) \citep{iwaya2019mobile}.
For many organisations \textbf{PIAs and privacy are seen as a competitive advantage} \citep{bayley2012privacy}, differentiating them from competitors through higher levels of privacy and data protection \citep{parks2011understanding} and demonstrating that their products are safe \citep{vandercruysse2020typology}.

\subsubsection*{N1 - Methodological Weaknesses}
\label{sec:n1}
\begin{landscape}
\begin{figure}[ht]
  \centering
  \includegraphics[width=\linewidth]{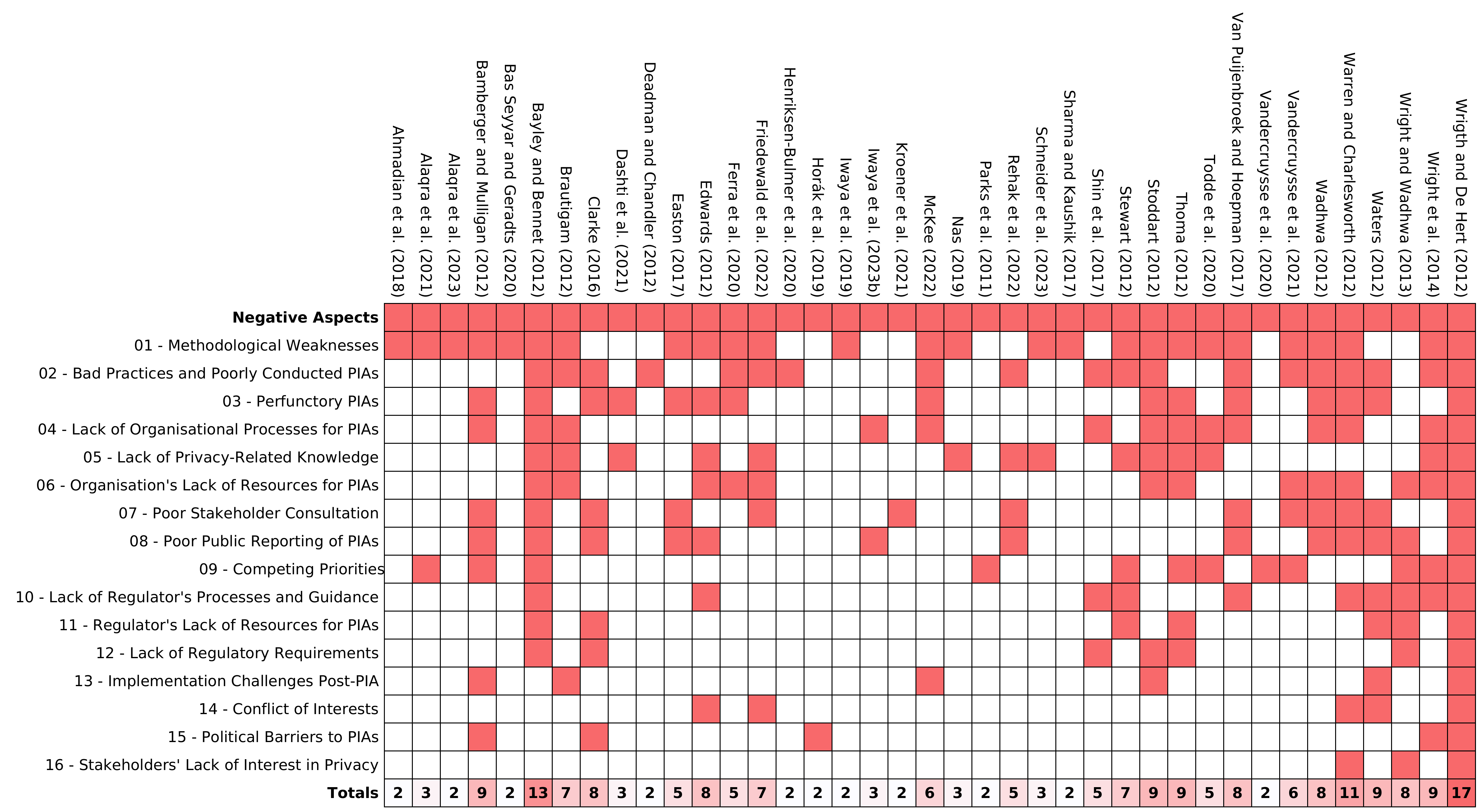}
  \caption{Overview of studies distributed across the sub-themes of Negatives Aspects related to PIAs in practice.}
  \label{fig:negative-distribution}
\end{figure}
\end{landscape}

Moving to the negative aspects of PIAs in practice, the identified sub-themes are distributed over a total of 39 studies, as shown in Figure \ref{fig:negative-distribution}.
Several studies stressed the \textbf{inherent challenges of risk assessments} that also apply to PIA methodologies.
Risk assessments elicit privacy threats and calculate the likelihood and severity of risks, which are based on the description of the system and the best knowledge of practitioners. In turn, this creates a shifting ground for PIAs, especially when assessing new and innovative technologies that may be used for unforeseen purposes and circumstances \citep{easton2017analysing}. As discussed in \citet{cherdantseva2016review} and \citet{iwaya2019mobile}, the estimation of risk in a mathematical sense is never complete, every possible undesired event (threat) is never known, and countermeasures for new attacks and vulnerabilities may not exist yet, so assessments always need to be continuously revised.

Another barrier that practitioners express is that \textbf{PIA methodologies need to be streamlined or are too complicated and complex}.
As reported in \citet{wright2014integrating}, many consider PIA frameworks ``not business friendly,'' ``too onerous,'' or ``too dense and complex,'' deeming them unsuitable to support the objectives of the company.
Other studies proposing new PIA methodologies also faced criticism, in which industry experts perceived the methodologies as trying to do too much and no longer being simple and easy to follow \citep{mckee2022pia}.

Some studies also criticised the \textbf{low credibility or proven effectiveness of PIA-related methods}, pointing to the limited evidence to indicate that PIAs are really attaining their stated goals in an efficient and cost-effective manner \citep{wadhwa2012privacy, friedewald2022dpias}.
Coupled with that, practitioners also see deficiencies in terms of the \textbf{subjectivity} of the risk analysis processes.
For many, the risk analysis is not only dependent on the skills and experience of the person doing the assessment, but deriving privacy risks (e.g., estimating likelihood and severity levels) is also perceived as a vague task \citep{van2017privacy}.

\subsubsection*{N2 - Bad Practices and Poorly Conducted PIAs}
\label{sec:n2}
Among the most common bad practices, the \textbf{inconsistent conduction of PIAs} was often reported, characterised by a lack of processes and clear responsibilities.
For example, the privacy officer's irregular involvement in the product and service development processes of the organisation would also lead to PIAs being conducted inconsistently \citep{deadman2012vodafone}.
Another major issue is the lack of planning with PIAs being carried out without a thought-out process \citep{vandercruysse2021dpia}.
Other organisations attempt to create their own PIA process but lack a systematic and comprehensive methodology that would enable them to fulfil regulatory requirements (e.g., GDPR) \citep{rehak2022analysis}.

The problem with \textbf{insufficient information for doing PIAs} was also reported, evidenced by the lack of documentation for the analysis \citep{brautigam2012pia}. The PIA preparation phase would often reveal such issues, showing how incomplete the knowledge of those in charge was when discussing the details and context of data processing operations \citep{friedewald2022dpias}.
Such problems would also result in \textbf{inappropriate risk analysis}, overlooking threats and risks, especially for scenarios beyond the normal processing activities, for instance when law enforcement gains access to data \citep{friedewald2022dpias}.
Another common issue in risk analysis is the \textbf{disregard for data subjects}, with organisations focusing on organisational risks instead of the risk to their customers or other individuals \citep{ferra2020challenges, henriksen-bulmer2020dpia}. This was also the case when dealing with risks to internal data subjects, for instance employees, seen as low priority \citep{friedewald2022dpias}.

\subsubsection*{N3 - Perfunctory PIAs}
\label{sec:n3}
PIAs are also reported to be conducted on a perfunctory basis \citep{wadhwa2012privacy}, that is, done merely out of duty with little interest or care.
The issue of \textbf{ceremonial or \textit{pro forma} performance of PIAs} has been linked to external mandates, where organisations adopt PIAs to appear legitimate but actually minimising the changes of existing practices and priorities \citep{bamberger2012pia}.
Practitioners have also reported that PIAs are often completed as a standard practice and then ``put on a shelf'' just to be referenced as a metric \citep{ferra2020challenges}.

Reports further show that \textbf{PIAs were conducted \textit{ex post facto} or when it was already too late}, and changes were limited and costly \citep{stoddart2012auditing}.
Late PIAs have marginal utility once the design parameters of the project have been set, an organisational structure is committed, and major costs incurred \citep{waters2012privacy}.
Practitioners have criticised such approaches, considering that organisations misunderstand the purpose of PIAs or might seek to mischaracterise them \citep{warren2008privacy}. 

\subsubsection*{N4 - Lack of Organisational Processes for PIAs}
\label{sec:n4}
Aspects such as the \textbf{lack of management support and infrastructure for PIAs} are considered barriers. The organisational process for managing PIAs needs to be defined and implemented, supported by specific actors in charge of maintaining and updating the documentation and produced assessments \citep{todde2020methodology}.
Furthermore, creating and sustaining such PIA infrastructures will be difficult without strong senior management support \citep{bayley2012privacy, stoddart2012auditing}.
The lack of a holistic privacy assessment process has also led many practitioners to report that \textbf{PIAs are not being carried out} in their organisations \citep{mckee2022pia}.
And when PIAs are conducted, it is common to see a \textbf{lack of internal audit evaluation processes}, with little or no auditing of the PIAs performed in practice \citep{van2017privacy}.

\subsubsection*{N5 - Lack of Privacy-Related Knowledge}
\label{sec:n5}
Practitioners mentioned the lack of privacy-related knowledge as one of the obstacles to implementing PIA practices.
For instance, \textbf{confusing privacy with security and confidentiality} and failing to recognise the boundaries between concepts would result in misjudgements about the necessity of a PIA or privacy impacts to data subjects \citep{dashti2021can}.
Another example is the \textbf{confusion or uncertainty about the roles of data processors and joint data controllers}, especially for organisations that operate in an ecosystem with several global service providers and 3rd parties \citep{nas2019data}.
The lack of knowledge about PIAs also requires practitioners to \textbf{clearly explain and set expectations to clients} regularly, clarifying what the assessment is intended to do and is capable of delivering \citep{edwards2012privacy}.

\subsubsection*{N6 - Organisation's Lack of Resources for PIAs}
\label{sec:n6}
Sometimes, organisations pointed to a lack of various types of resources needed for conducting PIAs. The \textbf{lack of PIA expertise or experts} was among the greatest difficulties encountered by some organisations, finding it hard to locate experienced practitioners outside the organisation \citep{warren2012privacy}, let alone identify internal people with such knowledge.
Practitioners also pointed to an alleged \textbf{lack of resources for conducting PIAs}, with typical arguments such as missing resources, expected delays, additional costs, lack of need, and reluctance to start the process \citep{thoma2012siemens}.
To some extent, even some DPAs have acknowledged that \textbf{PIAs are seen as a burden or a hurdle} and that the costs may be prohibitive for companies and agencies that do not have the financial capacity, incurring in seemingly unnecessary costs \citep{wadhwa2012privacy}.

\subsubsection*{N7 - Poor Stakeholder Consultation}
\label{sec:n7}
A challenge to the proper conduct of PIAs refers to the \textbf{lack of mechanisms for public participation or consultation}, limiting opportunities for outside experts to assist in assessing complex technological systems \citep{bamberger2012pia}.
This issue has also been linked to a methodological deficiency of popular PIA frameworks that lack concrete guidance or sufficiently describe citizen involvement processes \citep{vandercruysse2021dpia}.
Practitioners wrote that sometimes \textbf{public participation is obstructed and suggestions are ignored}, such as precluding the participation of advocacy groups or limiting comments to a rather late stage in the project, for instance after legislation is drafted and tabled \citep{clarke2016pia}.

\subsubsection*{N8 - Poor Public Reporting of PIAs}
\label{sec:n8}
A negative aspect of transparency is the poor public reporting of PIAs, both from the public sector and from the private sector.
For governmental institutions, \textbf{PIAs are not made available to the public}, or they are not easy to access, hidden on the websites of agencies without clear links; and for private companies, public PIAs are almost non-existent \citep{wadhwa2012privacy}.
Organisations also reported fears and barriers to releasing public PIAs, including concerns about the sensitivity of data and systems and the obstruction of self-criticism since participants would feel more guarded if subjected to a public open process. \citep{easton2017analysing}.

\subsubsection*{N9 - Competing Priorities}
\label{sec:n9}
The clash with other organisational priorities was also a significant hindrance for PIAs in practice. Project budgets are affected since \textbf{PIAs can incur extra costs and delays}, putting especially start-ups, small- and medium-sized companies at a disadvantage \citep{wadhwa2012privacy}.
Some practitioners also emphasised that meaningful engagement with \textbf{PIAs may disrupt the internal workings of an organisation} \citep{bamberger2012pia}.
Although this may be the exact purpose of PIA mandates, that is, forcing substantive change in the face of privacy-invasive operations, the internal reorientation process can be challenging.
In general, stakeholders think that the central functions of a service are to be kept and that privacy considerations should not unnecessarily impede business goals \citep{vandercruysse2021dpia}.

\subsubsection*{N10 - Lack of Regulatory Processes and Guidance}
\label{sec:n10}
The absence of some regulatory processes was also discussed in the following studies.
Regulatory challenges such as the \textbf{lack of mechanisms in the government for accountability of the implementation of PIAs plans} and promised mitigation measures were criticised by industry consultants \citep{bayley2012privacy}.
Other practitioners notice that \textbf{there is no central registry of PIA reports}, and there are no requirements to share reports with regulators or publish them anywhere  \citep{edwards2012privacy}. This adds to the opacity of the PIA processes in practice, making it hard to understand the uptake and evaluate the effectiveness of PIAs by the public and private sectors.

\subsubsection*{N11 - Regulator's Lack of Resources for PIAs}
\label{sec:n11}
Similar to companies, there is a \textbf{lack of resources for PIA-related activities on the regulator's side}. In practice, regulators are unlikely to receive enough resources for in-depth audits or to adequately supervise organisations and directly engage with several PIA project proponents \citep{waters2012privacy}.
Adding to the problem, changes in government and political disputes have also been used to \textbf{undermine regulatory agencies} by denying funding or obstructing their work \citep{clarke2016pia}.

\subsubsection*{N12 - Lack of Regulatory Requirements}
\label{sec:n12}
Given that \textbf{PIAs were not mandatory for the public or private sector} in many jurisdictions, institutions often would not conduct them \citep{thoma2012siemens}.
This lack of an explicit legal requirement and unified form and method leaves PIAs to be conducted based on the controller's discretion and goodwill, which is not always present when it comes to respecting privacy rights \citep{wright2013introducing}.
Although the EU has made PIAs mandatory when processing operations are \textit{``likely to result in a high risk to the rights and freedoms of natural persons,''} as stated in Article 35 of the GDPR \citep{GDPR2016}, in many countries, PIAs are still not a legal obligation but only ``encouraged'' \citep{clarke2011evaluation}.

\subsubsection*{N13 - Implementation Challenges Post-PIA}
\label{sec:n13}
Practitioners have highlighted \textbf{the challenge of implementing the identified measures after completing the PIA}. A core challenge is to have a supporting structure to follow up on the implementation status and to make the right people aware of the PIA findings so they can be truly implemented \citep{brautigam2012pia}.
Some reports further discuss when \textbf{PIAs do not bring on actual change in organisational practices}, for instance when clients do not welcome many recommendations, and there is significant tension in the finalisation of the report to mask core issues in the text so that potential critics cannot use the PIA work as effectively \citep{waters2012privacy}.

\subsubsection*{N14 - Conflict of Interests}
\label{sec:n14}
A critical factor that can compromise the credibility of PIAs is the \textbf{conflict of interests of internal or external assessors}.
Practitioners have discussed the challenge of achieving real independence in the PIA process since both in-house staff and consultants doing the PIAs have their livelihood depending on delivering work that pleases rather than annoys those who pay them \citep{edwards2012privacy}.
The organisation contracts most PIA practitioners and therefore expects them to act in the best interest of the business, making it difficult for them to assess risks in a truly unbiased manner, prioritising the concerns of the data subject \citep{friedewald2022dpias}.

\subsubsection*{N15 - Political Barriers for PIAs}
\label{sec:n15}
Some political barriers have been more particularly linked to the practice of PIAs within government agencies and in policy-making.
As discussed in \citet{bamberger2012pia}, information abuse in governmental initiatives is also a concern; however, expending political capital on privacy can be considered risky. Furthermore, there is also a clash among the values of \textbf{privacy versus efficiency and national security}, in which privacy can be seen as a hindrance for the others \citep{bamberger2012pia}.
Practitioners also reported other reasons for the \textbf{conflicts of PIAs in policy-making}, such as the complexity of this type of work, the involvement of several ministers and bargaining consultations, and the need ``to get on'' with the policy-making and final deliverables \citep{wright2014integrating}.

\subsubsection*{N16 - Stakeholders' Lack of Interest in Privacy}
\label{sec:n16}
Although stakeholder consultation is an essential part of the PIA process, there are significant challenges in engaging stakeholders and getting them involved.
As reported in \citet{warren2012privacy}, some PIA initiatives have faced a \textbf{lack of interest from stakeholders in participating in the consultation process}; despite sending several invitations to civil society representatives, very few took part in the consultation event and gave little to no meaningful feedback.

\subsubsection*{N17 - Dealing with the PIA Clients or Sponsors}
\label{sec:n17}
Handling the \textbf{disagreements on privacy risks between assessors and clients} is another significant challenge in the PIA process. Without compromising the independence of the report, assessors have to analyse trade-offs and be able to judge when to concede and when to make assertive PIA recommendations, knowing full well that they may be accepted or rejected \citep{edwards2012privacy}.

\subsection{The State of Existing Primary Research}
\label{sec:state-primary-research}
\subsubsection{Types of Primary Research}
\begin{figure}[htbp]
  \centering
  \includegraphics[width=0.5\linewidth]{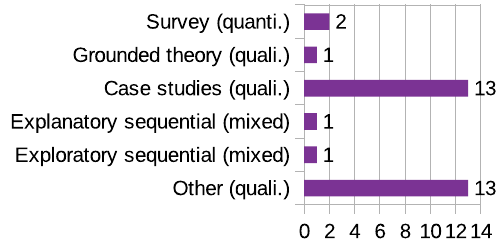}
  \caption{Types of Primary Research.}
  \label{fig:research-types-primary}
\end{figure}

This section addresses RQ3, delving further into the primary research on the topic.
As previously shown in Figure \ref{fig:research-types}, two-thirds of the studies were classified as Primary Research ($n=30$, 66.7\%).
These studies were further classified using \citet{creswell2017research} research design categories, as presented in Figure \ref{fig:research-types-primary}.

Twenty-eight studies (62.2\%) adopted a qualitative methodology, with case studies being the most common approach (28.9\%).
Thirteen studies (28.9\%) used other types of qualitative approaches in their research, including interview-based methods \citep{brautigam2012pia, bamberger2012pia, vandercruysse2020typology, pribadi2017regulatory, mckee2022pia, alaqra2021machine, schneider2023persona}, focus groups \citep{ferra2020challenges, vandercruysse2020typology, alaqra2023transparency, schneider2023persona, zamorano2022privacy}, Delphi study \citep{diiorio2009pia}, descriptive field study \citep{van2017privacy}, and cognitive walk-through \citep{alaqra2023transparency}.

Only four studies (10\%) used survey instruments as part of their research methods \citep{wright2014integrating, wright2013introducing, dashti2021can, vandercruysse2021dpia}.
Two studies were found to use a mixed-methods approach, being categorised as Explanatory Sequential \citep{wright2014integrating} and Exploratory Sequential \citep{vandercruysse2021dpia}, which relied on both quantitative surveys and qualitative interview-based methods.
Nonetheless, we argue that there is still an overall lack of primary research on the topic of PIAs in practice, especially in terms of quantitative research.





\subsubsection{Critical Appraisal of Qualitative Studies}
\label{sec:critical-quali}
As summarised in Tables \ref{tab:quali-appraisal} and \ref{tab:quanti-appraisal}, the quality of the identified primary research studies was critically appraised using the checklists from the Center for Evidence-Based Management \citep{cebma2014cap-quali, cebma2014cap-surveys}. Table \ref{tab:quali-appraisal} shows that the vast majority of the qualitative studies clearly defined a research question or objective (Q1) coupled with the use of appropriate methodologies (Q2). The research context was also sufficiently described in almost all studies (Q3). 

The main issues for most studies concerned the descriptions of the data collection (Q4-Q5) and the data analysis procedures (Q6-Q7). The data collection was often described briefly, without detailing the number of participants involved, time spent in interviews, procedures for recording and transcribing data, and missing interview guides or lists of questions used. In terms of allowing the data to be inspected (Q5), it was predictable that most studies would be unable to make their datasets available since they must maintain the participants' anonymity and avoid confidentiality breaches. Procedures for data analysis were also lacking since most studies did not describe the data analysis approach (e.g., grounded theory, thematic analysis), the number of researchers involved in the analysis, and other quality control measures. Regarding the repetition of the analysis steps, only the study of \citet{diiorio2009pia} was considered adequate since the authors use a Delphi panel for generating consensus, thus inherently seeking reliable agreement. 

Nonetheless, most studies were deemed to present credible results (Q8) and justified conclusions (Q9). However, it could be argued that the deficiencies of the studies in terms of the descriptions of data collection and analysis can compromise the final credibility of the results. Finally, most studies still present findings that are transferable to other settings (Q10) in the sense that readers can connect elements of the findings, consider and compare them to their own experience or similar situations, and infer that similar results could be achieved.

\begin{landscape}
\begin{table}[htbp]
\footnotesize
\begin{center}
\caption{Critical Appraisal of Qualitative Research --  Checklist for Qualitative Studies \citep{cebma2014cap-quali}. A full list of Questions is found in \ref{sec:quali-checklist}.}
\begin{tabular}{|lccccccccccc|}
\hline
\textbf{Author, year and refer.} & \textbf{Res. Design} & \textbf{Q1} & \textbf{Q2} & \textbf{Q3} & \textbf{Q4} & \textbf{Q5} & \textbf{Q6} & \textbf{Q7} & \textbf{Q8} & \textbf{Q9} & \textbf{Q10} \\ \hline
\hline
\citet{horak2019gdpr} & Case Study & Yes	& Yes	& Yes	& No	& No	& No	& No	& Yes	& Yes	& Yes \\
\citet{ferra2020challenges} & Other & Yes	& Yes	& Yes	& Yes	& No	& No	& No	& Yes	& Yes	& Yes \\
\citet{brautigam2012pia} & Other & Yes	& No	& Yes	& No	& No	& No	& No	& Yes	& Yes	& Yes \\
\citet{bamberger2012pia} & Other & Yes	& No	& Yes	& No	& No	& No	& No	& Yes	& Yes	& Yes \\
\citet{ahmadian2018supporting} & Case Study & Yes	& Yes	& Yes	& No	& No	& No	& No	& Yes	& Yes	& Yes \\
\citet{diiorio2009pia} & Other & Yes	& Yes	& Yes	& Yes	& No	& Yes	& Yes	& Yes	& Yes	& Yes \\
\citet{vandercruysse2020typology} & Case Study, Other & Yes	& Yes	& Yes	& Yes	& No	& No	& No	& Yes	& Yes	& Yes \\
\citet{iwaya2019mobile} & Case Study & Yes	& Yes	& Yes	& No	& No	& No	& No	& Yes	& Yes	& Yes \\
\citet{wright2014integrating} & Explan. Seq. & Yes	& Yes	& Yes	& No	& No	& No	& No	& Yes	& Yes	& Yes \\
\citet{clarke2016pia} & Case Study & Yes	& Yes	& Yes	& Yes	& Yes	& No	& No	& Yes	& Yes	& Yes \\
\citet{todde2020methodology} & Case Study & Yes	& Yes	& No	& No	& No	& No	& No	& No	& No	& Yes \\
\citet{parks2011understanding} & Grounded Theory & Yes	& Yes	& Yes	& Yes	& No	& Yes	& No	& Yes	& Yes	& Yes \\
\citet{pribadi2017regulatory} & Other & Yes	& Yes	& Yes	& No	& No	& No	& No	& Yes	& Yes	& Yes \\
\citet{sharma2017strategy} & Case Study & Yes	& Yes	& Yes	& No	& No	& No	& No	& Yes	& Yes	& Yes \\
\citet{warren2012privacy} & Case Study & Yes	& No	& Yes	& No	& No	& No	& No	& Yes	& Yes	& Yes \\
\citet{henriksen-bulmer2020dpia} & Case Study & Yes	& Yes	& Yes	& Yes	& No	& Yes	& No	& Yes	& Yes	& Yes \\
\citet{basseyyar2020pia} & Case Study & Yes	& Yes	& Yes	& No	& No	& No	& No	& Yes	& Yes	& Yes \\
\citet{van2017privacy} & Other & Yes	& Yes	& Yes	& No	& No	& No	& No	& Yes	& Yes	& Yes \\
\citet{kroener2021agile} & Case Study & Yes	& Yes	& Yes	& No	& No	& No	& No	& Yes	& Yes	& Yes \\
\citet{campanile2022evaluating} & Case Study & Yes	& Yes	& Yes	& No	& No	& No	& No	& Yes	& Yes	& Yes \\
\citet{friedewald2022dpias} & Case Study & Yes	& Yes	& Yes	& Yes	& No	& No	& No	& Yes	& Yes	& Yes \\
\citet{vandercruysse2021dpia} & Explor. Seq. & Yes	& Yes	& Yes	& Yes	& Yes	& No	& No	& Yes	& Yes	& Yes \\
\citet{mckee2022pia} & Other & Yes	& Yes    & Yes	& No	& No	& No	& No	& Yes	& No	& Yes \\
\citet{alaqra2023transparency} & Other & Yes	& Yes	& Yes	& No	& No	& No	& No	& Yes	& Yes	& Yes \\
\citet{zamorano2022privacy} & Other & Yes	& Yes	& Yes	& No	& No	& No	& No	& Yes	& Yes	& Yes \\
\citet{iwaya2023mental} & Other & Yes	& Yes	& Yes	& Yes	& No	& Yes	& Yes	& Yes	& Yes	& Yes \\
\citet{schneider2023persona} & Other & Yes	& Yes	& Yes	& No	& No	& No	& No	& Yes	& Yes	& Yes \\
\citet{alaqra2021machine} & Other & Yes	& Yes	& Yes	& Yes	& No	& Yes	& Yes	& Yes	& Yes	& Yes \\
\hline
\end{tabular}
\label{tab:quali-appraisal}
\end{center}
\end{table}
\end{landscape}

\subsubsection{Critical Appraisal of Quantitative Studies}
\label{sec:critical-quanti}
Considering the results from Table \ref{tab:quanti-appraisal}, it was found that all studies had focused research questions/objectives (Q1), employed appropriate methods for addressing them (Q2), and sufficiently explained the research context (Q3). Most issues for quantitative studies come from the possible bias when selecting participants (i.e., most just used convenience samples) (Q4), the lack of representativeness of the samples (Q5) and size for statistical power (Q6). The response rate was usually not discussed in the studies (Q7). The studies also did not describe steps for ensuring the reliability and validity of survey instruments (Q8). Statistical analysis was only assessed in one of the studies \citep{dashti2021can} (Q9), but none of the studies presented confidence intervals (Q10). Most studies also did not describe potential confounding variables (Q11). 

In conclusion, most results from the studies cannot be applied to other organisations in the strong sense of generalisability. The study of \citet{wright2013introducing} was an exceptional case since the authors conducted a survey study with the Data Protection Authorities (DPAs) of the 27 EU member states. The study sample can be considered representative since 17 out of the 27 DPAs in the EU responded to the survey, indicating an excellent response rate of 63\%. This survey \citep{wright2013introducing} was mostly an opinion-based questionnaire (i.e., free-form text responses), so statistical analysis would not apply, yet for some questions, the authors also organised the answers in categories with percentages. Nonetheless, we still consider that, in general, the methodological appropriateness of the quantitative studies lacks sufficient rigour.

\begin{landscape}
\begin{table}[htbp]
\footnotesize
\begin{center}
\caption{Critical Appraisal of Quantitative Research -- Checklist for Cross-Sectional Studies (Surveys) \citep{cebma2014cap-surveys}. A full list of Questions is found in \ref{sec:quanti-checklist}.}
\begin{tabular}{|lccccccccccccc|}
\hline
\textbf{Author, year and refer.} & \textbf{Res. Design} & \textbf{Q1} & \textbf{Q2} & \textbf{Q3} & \textbf{Q4} & \textbf{Q5} & \textbf{Q6} & \textbf{Q7} & \textbf{Q8} & \textbf{Q9} & \textbf{Q10} & \textbf{Q11} & \textbf{Q12} \\
\hline
\hline
\citet{wright2014integrating} & Expla. Seq. & Yes	& Yes	& Yes	& Yes	& No	& No	& No	& No	& No	& No	& Yes	& No \\
\citet{wright2013introducing} & Survey & Yes	& Yes	& Yes	& Yes	& Yes	& n.a.	& Yes	& n.a.	& n.a.	& n.a.	& Yes	& Yes \\
\citet{dashti2021can} & Survey & Yes	& Yes	& Yes	& Yes	& No	& No	& No	& No	& Yes	& No	& Yes	& No \\
\citet{vandercruysse2021dpia} & Explo. Seq. & Yes	& Yes	& Yes	& Yes	& No	& No	& No	& No	& No	& No	& Yes	& No \\
\hline
\end{tabular}
\label{tab:quanti-appraisal}
\end{center}
\end{table}
\end{landscape}

\section{Discussion}
\label{sec:discussion}
This ScR provides a comprehensive overview of the topic of PIAs in the wild, synthesising the literature and research in the area. In the following subsections, the main findings, challenges, and pathways for future research are further discussed, and the implications for research and practice are also presented.

\subsection{PIAs Before and After GDPR}
\begin{figure}[htbp]
  \centering
  \includegraphics[width=0.95\linewidth]{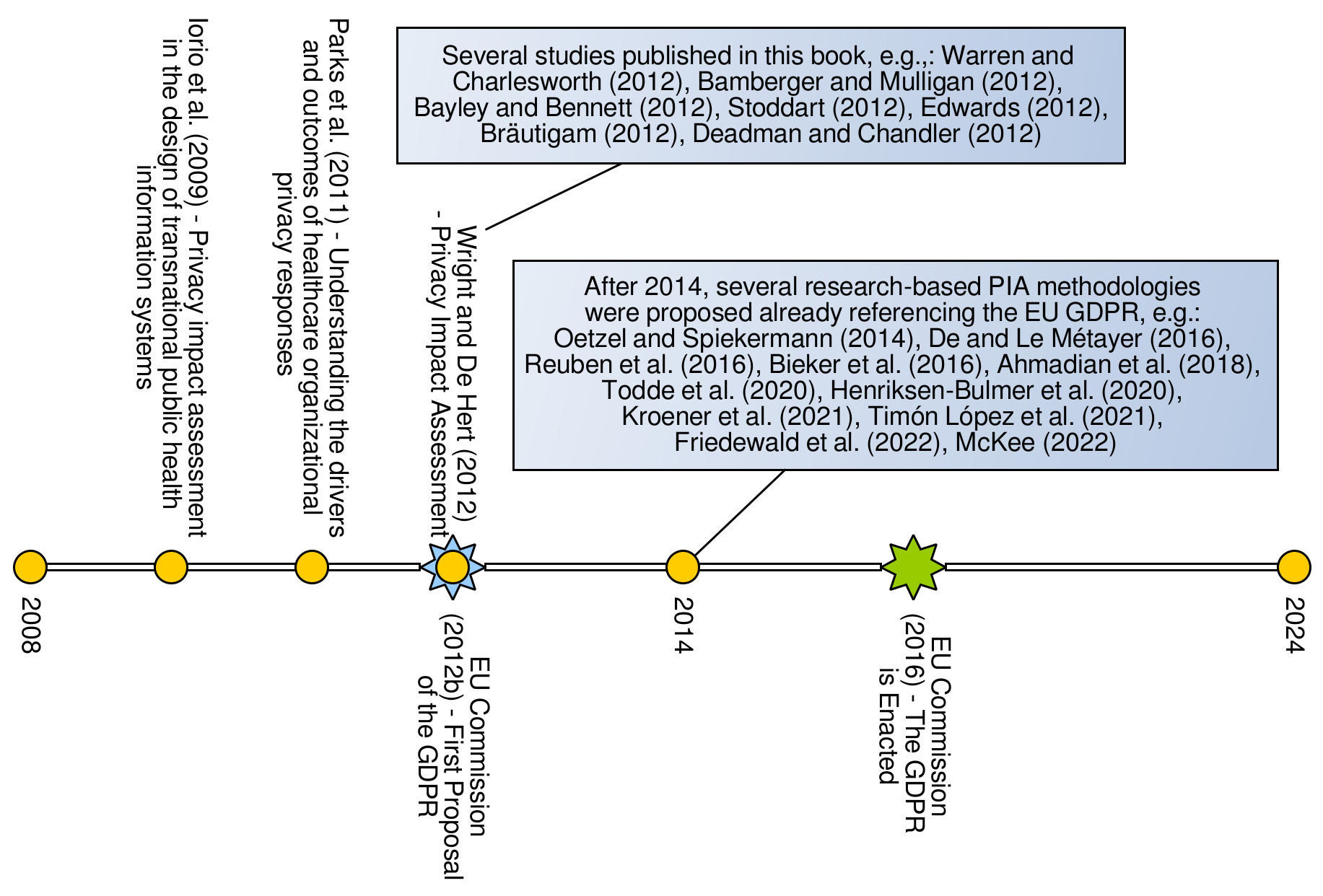}
  \caption{Timeline overview of the evolution of research on PIAs pre- and post-GDPR.}
  \label{fig:timeline-summary}
\end{figure}

Although several countries today have enforced privacy laws, the European GDPR \citep{GDPR2016} is undoubtedly the legal framework that has motivated most of the existing research on PIA frameworks and methodologies.
As previously shown in Figure \ref{fig:regulations}, 27 out of the 45 studies in this ScR have mentioned the GDPR in their text. In fact, a parallel can be drawn between the timeline of the GDPR and the evolution of PIA methodologies, as reflected in the included studies (see Figure \ref{fig:timeline-summary}). Before 2012, only a few studies had empirically investigated PIAs in practice, such as the works of \citet{diiorio2009pia} and \citet{parks2011understanding}. At this earlier stage, other studies were also primarily pointing to governmental PIA guidelines (e.g., in the UK \citep{warren2012privacy}, the US \citep{bamberger2012pia}, Canada \citep{bayley2012privacy, stoddart2012auditing}, and New Zealand \citep{edwards2012privacy}) or provided just a quick glance at industry approaches to PIAs (e.g., Nokia \citep{brautigam2012pia} and Vodafone \citep{deadman2012vodafone}).

It was in 2012 that the first proposal of the GDPR was released, which included obligations to data controllers in terms of conducting DPIAs \citep{GDPR2012}. During 2011 and 2013, more PIA frameworks and methodologies emerged, such as the RFID PIA \citep{bsi2011piarfid}, employing more systematic risk-based methods for identifying threats, associating severity and likelihood levels, and assigning controls \citep{oetzel2013systematic}. However, most studies ($n=28$, 62.2\%) included in this ScR were published after 2017 (see Figure \ref{fig:pubs-year}), when the GDPR had already been enacted. Since then, several research-based PIA methodologies \citep{de2016priam, reuben2016privacy, bieker2016process, ahmadian2018supporting, todde2020methodology, henriksen-bulmer2020dpia, kroener2021agile, timon2021approaching, friedewald2022dpias, mckee2022pia} and standardisation initiatives \citep{ISO29134} have been proposed. Nonetheless, advances in developing full-fledged PIA methodologies are still recent, occurring mainly in the past decade.

Considering only the primary research identified in the ScR ($n=30$), 22 studies (73.3\%) published between 2014 and 2023 primarily refer to the GDPR as the motivating legal framework for their research on PIAs. The remaining eight studies were all published before 2017, with three of them referring to the EU Data Protection Directive instead \citep{brautigam2012pia, diiorio2009pia, warren2012privacy} and five studies concentrated on other privacy regulations \citep{bamberger2012pia, clarke2016pia, parks2011understanding, pribadi2017regulatory, sharma2017strategy}. Such findings go to show the expressive impact of the GDPR in the development of PIA methodologies and the existing research on PIAs in practice. On the other hand, this predominance of GDPR-based PIA research may also make it challenging to transfer results to other countries with different privacy laws and PIA requirements. Therefore, non-EU practitioners, researchers, and policymakers should carefully consider existing findings to see if they also apply to their context.

On the other hand, the GDPR has clarified important practical aspects of how PIAs (i.e., DPIAs) should be conducted in an EU context, with such aspects directly connected to our positive and negative subthemes. For instance, issues related to accountability and who is responsible for the implementation of PIAs (Section \ref{sec:n10}.N10) were clarified with the entity of a ``data controller,'' bearing the responsibility, and the DPO, obligated to monitor and ensure compliance with the law (Sections \ref{sec:p1}.P1 and \ref{sec:p11}.P11). The lack of regulatory requirements for PIAs (Section \ref{sec:n11}.N11) was also addressed in Article 35 of the GDPR \citep{GDPR2016}, making it mandatory for high-risk systems that process personal data (Section \ref{sec:p1}.P1). A PIA must also be prepared \emph{before} beginning any data processing activity (Section \ref{sec:p2}.P2). Moreover, some issues around stakeholder consultation during the PIA process (Section \ref{sec:n7}.N7) have also been addressed, highlighting the DPO's role and key stakeholders' engagement in the project (Sections \ref{sec:p4}.P4 and \ref{sec:p6}.P6).

Furthermore, among the included studies in the ScR, it is clear that \cite{wright2012privacy} remains the most influential piece of work on PIAs in practice. The book was published in the same year that the first proposal of the GDPR was released, compiling knowledge from several privacy practitioners and academics worldwide. To a great extent, the work of \citet{wright2012privacy}, albeit conducted pre-GDPR, laid the foundation for PIA research in the following years. On the other hand, prominent research has also been published post-GDPR, particularly the works of \citet{ahmadian2018supporting}, \citet{horak2019gdpr}, and \citet{vandercruysse2020typology}, which have been highly cited, presenting important case studies of PIAs in real-world projects, heavily motivated by the challenges of complying with the GDPR legal requirements.


\subsection{Overarching Research Gaps}
As evidenced in this ScR, there is an overall lack of literature and research on PIAs in practice, giving rise to different \emph{types of research gaps}, as proposed in \citet{miles2017taxonomy}. Apart from the limited number of identified studies, many of the publications are book chapters (i.e., $n=13$, 32.5\%), so their findings can be questioned due to the absence of formal scientific peer review. This leaves only 32 peer-reviewed studies (i.e., 17 journal articles, 14 conference papers, and one doctoral thesis). In what follows, future research pathways are introduced according to the four main types of research gaps identified in this ScR, that is, evidence-related, methodological, empirical, and population-related gaps \citep{miles2017taxonomy}.

\subsubsection{Evidence Gap}
There is an evidence gap since many studies yield contradictory results, such as disputing the proven effectiveness of PIA methodologies (Section \ref{sec:n1}.N1) \citep{wadhwa2012privacy, friedewald2022dpias}, whether or not PIAs should be mandatory (Section \ref{sec:n6}.N6) \citep{wright2013introducing}, how to streamline PIA approaches (Section \ref{sec:n1}.N1) \citep{wright2013introducing} yet maintaining the assessment comprehensive and rigorous (Section \ref{sec:n2}.N2) \citep{clarke2016pia}, and the actual feasibility of conducting truly independent PIAs (Section \ref{sec:n14}.N14) \citep{edwards2012privacy, friedewald2022dpias}. Finding the balance and achieving consensus on such issues will require further research and discussion among all stakeholders; a one-size-fits-all PIA approach is unlikely.

\subsubsection{Methodological Gap}
This ScR also points to significant methodological gaps, meaning that adequate study designs for the rigorous evaluation of PIA methodologies via empirical inquiry still need to be improved (Section \ref{sec:n1}.N1) \citep{wadhwa2012privacy, friedewald2022dpias}. Furthermore, the critical appraisals for existing studies (Sections \ref{sec:critical-quali} and \ref{sec:critical-quanti}) point to significant deficiencies in terms of methodological quality, especially concerning the procedures for data collection and analysis. It is also worth noticing that most evidence comes from \emph{Case Study} and \emph{Expert Opinion} publications, which are, regrettably, still at lower levels of evidence strength if compared to other types of study design (e.g., controlled studies, cohort studies, and randomised control trials). The topic of PIAs is currently ripe for further and more sophisticated research designs towards building better evidence from practice. 


\subsubsection{Empirical Gap}
Although much research has been published on the broad topic of PIAs and DPIAs, this ScR shows that there is still a lack of empirical studies in real-world practice. This can be considered a major empirical gap concerning the evaluation and practical verification of many other PIA methodologies that have been proposed in recent years. This ScR has identified some prominent studies that help to fill this empirical gap (e.g., \citet{diiorio2009pia, parks2011understanding, friedewald2022dpias, henriksen-bulmer2020dpia, vandercruysse2021dpia}). However, many other PIA methodologies could still be further evaluated, approaches compared, and studies reproduced, looking to corroborate or add new findings.

This issue concerning the lack of evaluation research on PIAs has also been raised by other researchers (see for instance \citet{friedewald2022dpias}). This lack of evaluation research was also found in the area of privacy engineering in general \citep{gurses2016priveng, iwaya2023privacy}. For this reason, we argue that researchers and practitioners could focus on developing future work on the topic of \emph{Empirical Privacy Engineering}. This includes PIAs/DPIAs and many other theories, methods, techniques, and tools (e.g., privacy threat modelling, privacy risk assessments, requirements engineering for privacy) in the emerging research area of privacy engineering that still need to be rigorously evaluated \citep{gurses2016priveng}.

\subsubsection{Population Gap}
Another significant challenge refers to the population gap revealed in the included studies, showing that PIA research also suffers from an over-reliance on Western, Educated, Industrialised, Rich, and Democratic (WEIRD) populations \citep{henrich2010most}, mainly from Europe and North America. This ScR has nonetheless identified a minority of studies ($n=6$, 15\%) addressing the practice of PIA in populations from Hong Kong \citep{waters2012privacy}, Indonesia \citep{pribadi2017regulatory}, Japan \citep{shin2017analysis}, India and the Philippines \citep{sharma2017strategy}, Brazil \citep{iwaya2019mobile}, Syria, Yemen and Iraq \citep{kroener2021agile}. For such reasons, the above-mentioned research gaps have implications for researchers, practitioners, and policymakers tasked with further developing evidence-based PIA practices.

\subsubsection{Further Systematic Reviews}
The comprehensive nature of scoping reviews, as explained in Section \ref{sec:methods}, also enables the identification of potential topics for more focused systemic literature reviews (SLRs). Since this ScR points to a growing maturity of research-based PIA methodologies, one potential topic for a future SLR could be the identification of existing PIA methodologies that have been empirically tested, even if not fully implemented and evaluated in practice. Future SLRs can also focus on how specific stakeholders, such as data subjects, data protection supervisory authorities, or policymakers, perceive and use PIAs. Similarly, other privacy engineering methodologies can be targeted in future systematic reviews, such as model-based approaches for privacy-friendly systems or privacy requirements engineering methods, as identified by \citet{alslais2020priveng}.

\subsection{Considerations for PIA Practitioners and Policymakers}
During this ScR, it was also perceived that some key considerations could be discussed, and recommendations could be drawn to better inform PIA practitioners and policymakers. In what follows, we discuss the risks of distorting PIA attempts and expand on the future requirements (e.g., AI Act) to assess sociotechnical systems from a broader human rights perspective.

\subsubsection{The Risk of Distorting PIAs}
A PIA aims at providing an accurate, well-founded, and reliable impact assessment. An erroneous, distorted, or misleading PIA may even create a risk because it should be a reliable basis for taking action, specifically deciding whether the data processing should be conducted as planned and selecting the appropriate technological and organisational measures for sufficiently mitigating the risks (Section \ref{sec:n2}.N2) \citep{brautigam2012pia, friedewald2022dpias}. Without a proper basis, the lawfulness of the processing and its compliance with legal provisions may be at stake, there may be undetected risks, and the set of selected measures may not be suitable for proper risk treatment. As a result, the costs may rise if later the processing has to be changed, and additional costs may occur due to fines or financial compensation for data subjects (Section \ref{sec:n3}.N3) \citep{stoddart2012auditing, waters2012privacy}.

There are distorted PIAs that are only commissioned work for marketing purposes or even for usage as fig leaf against criticism concerning the processing or the measures taken (Section \ref{sec:n3}.N3) \citep{bamberger2012pia,wadhwa2012privacy, warren2008privacy}. But even apart from those PIAs, it is a challenging endeavour to conduct a comprehensive PIA in a neutral way without prejudging the outcome. The team performing the PIA may be professionally blinkered or short-sighted concerning risks due to their experience and their will to reach a positive conclusion instead of identifying risks not yet considered. 

Moreover, some methodologies or tools lead to oversimplification or premature results of a PIA. In particular, this is the case if the PIA relies on a \textit{``checklist approach''} \citep{bieker2016process} (e.g., UK ICO PIA, CNIL PIA), where the PIA team may be tempted to only check off a given list of properties of the IT system without a real understanding of the risks. The same is valid for assessing potential technological and organisational measures \citep{bieker2016process}. However, it is also worth noting that even thorough PIA methodologies can be misused by practitioners due to their lack of privacy knowledge (Section \ref{sec:n5}.N5) \citep{dashti2021can, nas2019data} or due to the perfunctory conduction of PIAs (Section \ref{sec:n3}.N3) \citep{bamberger2012pia, wadhwa2012privacy}. 

The PIA should be comprehensible for other parties, including auditors and relevant stakeholders (which may also encompass the public) (Section \ref{sec:p4}.P4) \citep{deadman2012vodafone, friedewald2022dpias}. Here, it is necessary to decide on the appropriate levels of granularity and complexity \citep{martin2020dpiareport}. Again, a simple checklist approach would not be fit for this; instead, the performance of a PIA should be considered as an iterative task (Section \ref{sec:p2}.P2) \citep{basseyyar2020pia}. 

\subsubsection{From PIAs to Fundamental Rights Impact Assessments}
PIAs, and similarly DPIAs, usually focus on the protection of personal data or the privacy of individuals. Thus, the starting point is the (planned) processing of personal data. However, there are further risks to fair design, our society, our planet, and so on. Not all of these additional risks arise from processing personal data, and not all affect individuals, that is, the data subjects in the GDPR terminology. Recognising this limitation, there have been developments towards Human Rights Impact Assessment (HRIA) of digital activities \citep{danish2020hria} and Fundamental Rights Impact Assessment (FRIA) \citep{janssen2022practicalfria, dutch2020fraia}.

Recent European legislation, such as the Digital Services Act (DSA, Regulation (EU) 2022/2065) and the upcoming AI Act, demand FRIAs for high-risk data processing contexts. Among others, the risks to be assessed under the DSA regime encompass the dissemination of illegal content, negative effects on civic discourse and the electoral process, public security, and negative effects in relation to gender-based violence, public health protection, the protection of minors, and serious negative consequences to the physical and mental well-being of persons (Article~34 DSA) \citep{mantelero2023friadsa}. The approach of the AI Act proposal from the European Parliament requires deployers of AI systems to conduct an FRIA of a system in the specific context of use and particularly mentions risks of harm likely to impact marginalised persons or vulnerable groups as well as the reasonably foreseeable adverse impact on the environment \citep{ep2023aiact}.

PIAs and DPIAs may be understood as pieces of the bigger picture of an FRIA. The same holds for specific assessments, for instance concerning the impact on civic discourse or the environment. However, it has to be noted that the various impact assessments should not be conducted separately. Instead, a holistic approach is necessary to tackle the interdependencies concerning risks and planned measures to mitigate those risks. Similarly to PIAs, FRIAs would also need to be integrated into the organisational risk management (Section \ref{sec:p8}.P8) \citep{wright2014integrating, parks2011understanding}.

\section{Threats to Validity}
\label{sec:threats-validity}
\subsection{Threat I -- Limitations of the ScR plan}
The first threat relates to the planning of the ScR in terms of identifying the need and justification for this study. Here, we were concerned with identifying existing reviews (systematic and non-systematic) on the topic of PIAs ``in the wild.'' The initial searches did not reveal any review studies on the topic, as described in \ref{sec:identify-scr-need}, pointing to a significant gap in secondary research on PIAs and DPIAs. The planning phase of the ScR is also critical to outline the research questions and provide the basis for an objective investigation of the studies that are being reviewed. If the RQs are not explicitly stated or omit the key topics, the scoping review results can be flawed, overlooking the key information. To mitigate this threat, we outlined three RQs and objectives (Section \ref{sec:research-questions}) that aim to find answers about the frequency, types of research, positive and negative aspects, the strength of primary research, and research gaps. In summary, we seek to minimise any bias or limitations during the planning phase when defining the scope and objectives of this ScR. As a last step in the planning phase, the team finalised and cross-checked the study protocol to minimise the limitations of the ScR plan before proceeding to the subsequent phases.

\subsection{Threat II -- Credibility of the Literature Search Process}
Identifying and selecting the studies reviewed in the ScR are also significant processes to be observed. Selecting studies is a critical step; if any relevant papers are missed, the results of the ScR may be flawed. Therefore, we followed a two-step process (Section \ref{sec:eligibility-criteria}), referred to as (i) literature screening and (ii) complete reading of papers. This selection process was carried out independently by two reviewers. We also performed forward and backward snowballing, looking for references to other potentially relevant studies. Also, this ScR restricts the selection of publications to four scientific databases: Scopus, Web of Science, IEEE Xplore, ACM Digital Library, and the OpenGrey database for grey literature. Only these databases were used due to their high relevancy to computer science, privacy, and data protection, as well as to maintain a feasible search space. We essentially used only two broad search terms in this ScR (``privacy impact assessments'' and ``data protection impact assessments''), maximising the search results; no other constraining terms or filters were necessary since the number of results from this search strategy was still feasible for investigation by the reviewers.
This step-wise search process gives us confidence that we minimised limitations related to (i) excluding or overlooking relevant studies or (ii) including irrelevant studies that could impact the results and their reporting in the ScR.

\subsection{Threat III -- Potential Bias in ScR Reporting}
Some threats should also be considered regarding the potential bias in synthesising the data from the review and documenting the results. This means that if there are some limitations in the data synthesis, they directly impact the results of this ScR. 
Typical examples of such limitations could be a flawed research taxonomy, incorrect identification of research themes (e.g., positive and negative) and a mismatch of potential research gaps.
To minimise the bias in synthesising and reporting the results, we have created a data extraction form that uses well-known classification schemes, such as the ones proposed by \citet{wieringa2006requirements} and \citet{creswell2017research}.
Three researchers independently reviewed this data extraction form while revising the research protocol.
While one of the researchers led the data extraction step, two other authors helped by cross-checking the work throughout the process for consistency. Three authors were involved in the thematic analysis of the positive and negative aspects derived from the literature. During the thematic analysis, they actively worked on reviewing the names of codes for consistency and defining themes and sub-themes through a series of meetings.
Furthermore, this ScR also offers a complete replication package \citep{iwaya2024repo} that conveniently enables other researchers to reproduce or extend this review, as described in Section \ref{sec:writing-protocol}.

\section{Conclusion}
\label{sec:conclusion}
Even though PIAs have a long history of developments in the policy arena \citep{CLARKE2009123}, they must become genuinely evidence-based processes for industry practitioners. Whilst significant contributions have been made in the academic arena concerning PIAs, this ScR shows that the time is ripe for further research since the area is still evolving. In fact, PIAs can be seen as one of the most promising research pathways in the field of Empirical Privacy Engineering, that is, using empirical research methods to study and evaluate engineering approaches for integrating privacy in working systems, taking part in the broad area of empirical software engineering \citep{gueheneuc2019empirical, fernandez2019empirical}. PIAs are a widely known strategy for privacy by design, and many prominent PIA frameworks today still need to be rigorously and independently evaluated. However, developing realistic research designs is a challenge in itself. Furthermore, if the proven effectiveness of PIAs still lacks strong empirical evidence, broader concepts on human rights impact assessments \citep{de2009human} remain arguably even less understood. Nonetheless, based on the existing research, much can be learned, moving toward more evidence-driven PIAs and privacy engineering practices.

\section{CRediT authorship contribution statement}
\textbf{Leonardo Horn Iwaya:} Conceptualization, Data curation, Formal Analysis, Investigation, Methodology, Visualization, Writing - original draft, Writing - review \& editing.
\textbf{Ala Sarah Alaqra}: Conceptualization, Investigation, Formal analysis, Writing - review \& editing.
\textbf{Marit Hansen}: Formal analysis, Writing - original draft, Writing - review \& editing.
\textbf{Simone Fischer-H\"{u}bner}: Formal analysis, Writing - original draft, Writing - review \& editing.

\section{Declaration of Competing Interest}
The authors declare that they have no known competing financial interests or personal relationships that could have appeared to influence the work reported in this paper.

\section{Acknowledgement}
This work was supported in part by the Knowledge Foundation of Sweden (KKS), Region V\"{a}rmland (Sweden) through the DHINO project (Grant: RUN/220266), and Sweden's Innovation Agency (Vinnova) via the DigitalWell Arena project (Grant: 2018-03025). Moreover, Simone Fischer-H\"{u}bner's work was also partially supported by the Wallenberg AI, Autonomous Systems and Software Program (WASP) funded by the Knut and Alice Wallenberg Foundation.

\section{Data availability}
\label{sec:online-resources}
The replication package for this ScR is provided in a public repository \citep{iwaya2024repo} and the complete ScR Research Protocol \citep{iwaya2023protocol}.

\section{Vitae}
\textbf{Leonardo Horn Iwaya} is an Associate Senior Lecturer in the Department of Mathematics and Computer Science, Karlstad University, Sweden. He obtained a Ph.D. degree in Computer Science from Karlstad University, Sweden, in 2019. He currently works with the Privacy \& Security (PriSec) Research Group at Karlstad University, contributing to projects such as CyberSecurity4Europe, TRUEdig, SURPRISE, DHINO, and DigitalWell Arena. His research interests include privacy engineering, cybersecurity, human factors, mobile and ubiquitous health systems, and the privacy impacts of new technologies.

\textbf{Ala Sarah Alaqra} is an Associate Professor (Docent) in Information Systems at Karlstad University, Sweden. Having a background in Computer Science working within the areas of Privacy Enhancing Technologies (PETs) and expertise in Human-Computer Interaction, Ala Sarah has a broad experience in interdisciplinary privacy and security research projects. Their general research interest involves the use and impacts of technological trends on users, organisations, and societies, including usable privacy, usable PETs, human-centred design (including AI), and eHealth applied contexts.

\textbf{Marit Hansen} has been the State Data Protection Commissioner of Land Schleswig-Holstein and Chief of Unabh\"{a}ngiges Landeszentrum f\"{u}r Datenschutz (ULD) since 2015. Before being appointed Data Protection Commissioner, she had been Deputy Commissioner for seven years. Within ULD, she established the ``Privacy Technology Projects'' Division and the ``Innovation Centre Privacy \& Security.'' She was also a member of the Data Ethics Commission of the German Government. Since earning her diploma in Computer Science in 1995, she has been working on privacy and security aspects. Marit's focus is on ``data protection by design'' and ``data protection by default'' from both the technical and the legal perspectives. 

\textbf{Simone Fischer-H\"{u}bner} is a Professor at the Department of Mathematics and Computer Science, Karlstad University, Sweden, where she is leading the Privacy \& Security (PriSec) research group. Additionally, she is also a Guest Professor at the Department of Computer Science and Engineering at the Chalmers University of Technology, Sweden. She is specialised in privacy-enhancing technologies, cybersecurity and interdisciplinary privacy aspects, including usable privacy. She is a board member of the Swedish Data Protection Forum (Forum f\"{o}r Dataskydd) and an advisory board member of the Swedish Cybersecurity Council.

\appendix

\section{Critical Appraisal Checklists}
\subsection{Critical Appraisal of Qualitative Research}
\label{sec:quali-checklist}
The Checklist for Qualitative Studies from \citet{cebma2014cap-quali} is composed of the following questions (Answers: Yes, No, Can't tell):
\begin{itemize}
    \item \textbf{Q1.} Did the study address a clearly focused question/issue?
    \item \textbf{Q2.} Is the research method (study design) appropriate for answering the research question?
    \item \textbf{Q3.} Was the context clearly described?
    \item \textbf{Q4.} How was the fieldwork undertaken? Was it described in detail? Are the methods for collecting data clearly described?
    \item \textbf{Q5.} Could the evidence (fieldwork notes, interview transcripts, recordings, documentary analysis, etc.) be inspected independently by others?
    \item \textbf{Q6.} Are the procedures for data analysis reliable and theoretically justified? Are quality control measures used?
    \item \textbf{Q7.} Was the analysis repeated by more than one researcher to ensure reliability?
    \item \textbf{Q8.} Are the results credible, and if so, are they relevant for practice?
    \item \textbf{Q9.} Are the conclusions drawn justified by the results?
    \item \textbf{Q10.} Are the findings of the study transferable to other settings?
\end{itemize}

\subsection{Critical Appraisal of Quantitative Research}
\label{sec:quanti-checklist}
The Checklist for Cross-Sectional Studies (Surveys) from \citet{cebma2014cap-surveys} is composed of the following questions (Answers: Yes, No, Can't tell):
\begin{itemize}
    \item \textbf{Q1.} Did the study address a clearly focused question/issue?
    \item \textbf{Q2.} Is the research method (study design) appropriate for answering the research question?
    \item \textbf{Q3.} Is the method of selection of the subjects (employees, teams, divisions, organisations) clearly described?
    \item \textbf{Q4.} Could the way the sample was obtained introduce (selection) bias?
    \item \textbf{Q5.} Was the sample of subjects representative with regard to the population to which the findings will be referred?
    \item \textbf{Q6.} Was the sample size based on pre-study considerations of statistical power?
    \item \textbf{Q7.} Was a satisfactory response rate achieved?
    \item \textbf{Q8.} Are the measurements (questionnaires) likely to be valid and reliable?
    \item \textbf{Q9.} Was the statistical significance assessed?
    \item \textbf{Q10.} Are confidence intervals given for the main results?
    \item \textbf{Q11.} Could there be confounding factors that haven’t been accounted for?
    \item \textbf{Q12.} Can the results be applied to other organisations?
\end{itemize}

\section{List of papers in the review}
\label{app:list-of-papers}
All the studies included in the ScR are listed in Tables \ref{tab:list-of-papers}, \ref{tab:list-of-papers-cont1}, and \ref{tab:list-of-papers-cont2}.

\begin{landscape}
\begin{table}[htbp]
\footnotesize
\begin{center}
\caption{Complete list of papers included in the scoping review. \textbf{Init Search:} studies selected from the initial database searches. \textbf{FW Snowb (\textit{forward snowballing})}: studies that cited the initially selected papers. \textbf{BW Snowb (\textit{backward snowballing})}: studies selected in full-text readings (i.e., in references).}
\begin{tabular}{|p{0.225\linewidth}|p{0.625\linewidth}|c|}
\hline
\textbf{Ref.} & \textbf{Title} & \textbf{Added in} \\ \hline \hline
\citet{horak2019gdpr} & GDPR Compliance in Cybersecurity Software: A Case Study of DPIA in Information Sharing Platform & Init. Search \\
\citet{ferra2020challenges} & Challenges in assessing privacy impact: Tales from the front lines & Init. Search \\
\citet{brautigam2012pia} & PIA: Cornerstone of Privacy Compliance in Nokia & Init. Search \\
\citet{bamberger2012pia} & PIA Requirements and Privacy Decision-Making in US Government Agencies & Init. Search \\
\citet{ahmadian2018supporting} & Supporting privacy impact assessment by model-based privacy analysis & Init. Search \\
\citet{diiorio2009pia} & Privacy impact assessment in the design of transnational public health information systems: The BIRO project & Init. Search \\
\citet{vandercruysse2020typology} & A typology of Smart City services: The case of Data Protection Impact Assessment & Init. Search \\
\citet{iwaya2019mobile} & Mobile health systems for community-based primary care: identifying controls and mitigating privacy threats & Init. Search \\
\citet{wright2014integrating} & Integrating privacy impact assessment in risk management & Init. Search \\
\citet{clarke2016pia} & Privacy impact assessments as a control mechanism for Australian counter-terrorism initiatives & Init. Search \\
\citet{wright2013introducing} & Introducing a privacy impact assessment policy in the EU member states & Init. Search \\
\citet{todde2020methodology} & Methodology and workflow to perform the Data Protection Impact Assessment in healthcare information systems & Init. Search \\
\citet{parks2011understanding} & Understanding the drivers and outcomes of healthcare organizational privacy responses & Init. Search \\
\citet{pribadi2017regulatory} & Regulatory recommendations for IoT smart-health care services by using Privacy Impact Assessment (PIA) & Init. Search \\
\citet{sharma2017strategy} & Strategy for privacy assurance in offshoring arrangements & Init. Search \\
\hline
\end{tabular}
\label{tab:list-of-papers}
\\
\end{center}
\end{table}
\end{landscape}

\begin{landscape}
\begin{table}[htbp]
\footnotesize
\begin{center}
\caption{Complete list of papers included in the scoping review. (Continued)}
\begin{tabular}{|p{0.225\linewidth}|p{0.625\linewidth}|c|}
\hline
\textbf{Ref.} & \textbf{Title} & \textbf{Added in} \\ \hline \hline
\citet{warren2012privacy} & Privacy impact assessment in the UK & Init. Search \\ 
\citet{wright2012findings} & Findings and recommendations & Init. Search \\
\citet{henriksen-bulmer2020dpia} & DPIA in context: Applying DPIA to assess privacy risks of cyber physical systems & Init. Search \\
\citet{bayley2012privacy} & Privacy impact assessments in Canada & Init. Search \\
\citet{stoddart2012auditing} & Auditing privacy impact assessments: The Canadian experience & Init. Search \\
\citet{basseyyar2020pia} & Privacy impact assessment in large-scale digital forensic investigations & Init. Search \\
\citet{easton2017analysing} & Analysing the Role of Privacy Impact Assessments in Technological Development for Crisis Management & Init. Search \\
\citet{van2017privacy} & Privacy impact assessment in practice: The results of a descriptive field study in the Netherlands & Init. Search \\
\citet{waters2012privacy} & Privacy impact assessment - Great potential not often realised & Init. Search \\
\citet{edwards2012privacy} & Privacy impact assessment in New Zealand - A practitioner’s perspective & Init. Search \\
\citet{kroener2021agile} & Agile ethics: an iterative and flexible approach to assessing ethical, legal and social issues in the agile development of crisis management information systems & Init. Search \\
\citet{shin2017analysis} & Analysis of specific personal information protection assessment in the social security and tax number system of local governments in Japan & Init. Search \\
\citet{wadhwa2012privacy} & Privacy impact assessment reports: A report card & Init. Search \\
\citet{stewart2012privacy} & Privacy impact assessment: Optimising the regulator’s role & Init. Search \\
\citet{rehak2022processing} & The Processing goes far beyond ``the app''- Privacy issues of decentralized Digital Contact Tracing using the example of the German Corona-Warn-App & Init. Search \\
\citet{campanile2022evaluating} & Evaluating the Impact of Data Anonymization in a Machine Learning Application & Init. Search \\
\citet{rehak2022analysis} & Analysis and Constructive Criticism of the Official Data Protection Impact Assessment of the German Corona-Warn-App & Init. Search \\
\hline
\end{tabular}
\label{tab:list-of-papers-cont1}
\\
\end{center}
\end{table}
\end{landscape}

\begin{landscape}
\begin{table}[htbp]
\footnotesize
\begin{center}
\caption{Complete list of papers included in the scoping review. (Continued)}
\begin{tabular}{|p{0.225\linewidth}|p{0.625\linewidth}|c|}
\hline
\textbf{Ref.} & \textbf{Title} & \textbf{Added in} \\ \hline \hline
\citet{friedewald2022dpias} & Data Protection Impact Assessments in Practice: Experiences from Case Studies & Init. Search \\
\citet{rajamaki2021design} & Design Science Research Towards Ethical and Privacy-Friendly Maritime Surveillance ICT Systems & Init. Search \\
\citet{dashti2021can} & Can data subject perception of privacy risks be useful in a data protection impact assessment? & Init. Search \\
\citet{nas2019data} & Data protection impact assessment: Assessing the risks of using Microsoft Office ProPlus & Init. Search \\
\citet{thoma2012siemens} & How Siemens assesses privacy impacts & Init. Search \\
\citet{deadman2012vodafone} & Vodafone’s approach to privacy impact assessments & Init. Search \\
\citet{alaqra2023transparency} & Transparency of Privacy Risks Using PIA Visualizations & Init. Search \\
\citet{zamorano2022privacy} & Privacy by Design in CBRN Technologies Targeted to Vulnerable Groups: The Case of PROACTIVE & Init. Search \\
\citet{iwaya2023mental} & On the privacy of mental health apps: An empirical investigation and its implications for app development & Init. Search \\
\citet{schneider2023persona} & Persona-oriented Data Protection Impact Assessment for Small Businesses & Init. Search \\
\citet{alaqra2021machine} & Machine learning–based analysis of encrypted medical data in the cloud: Qualitative study of expert stakeholders’ perspectives & BW Snowb. \\
\citet{vandercruysse2021dpia} & The DPIA: Clashing Stakeholder Interests in the Smart City? & FW Snowb. \\
\citet{mckee2022pia} & Privacy Assessment Breakthrough: A Design Science Approach to Creating a Unified Methodology & FW Snowb. \\
\hline
\end{tabular}
\label{tab:list-of-papers-cont2}
\\
\end{center}
\end{table}
\end{landscape}

\bibliographystyle{elsarticle-harv} 
\bibliography{cas-refs}





\end{document}